\documentclass[fleqn,usenatbib]{mnras}

\usepackage{amsmath}
\usepackage{amssymb}
\usepackage{amsfonts}
\usepackage{siunitx}
\usepackage{xcolor}
\usepackage{journals}
\usepackage{orcidlink}
\usepackage{booktabs}
\usepackage{comment}
\usepackage{cleveref}
\crefname{table}{Tab.}{Tabs.}
\crefname{section}{App.}{Apps.}
\crefname{fig}{Fig.}{Figs.}
\newcommand{\orcid}[1]{\href{https://orcid.org/#1}{\includegraphics[width=10pt]{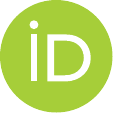}}}

\newcommand{\etal}{\emph{et al.}}
\newcommand{\dif}{\mathop{}\!\mathrm{d}}
\newcommand{\gperccc}{\ensuremath{\,\mathrm{g}\,\mathrm{cm}^{-3}}}
\newcommand{\dynpercc}{\ensuremath{\,\mathrm{dyn}\,\mathrm{cm}^{-2}}}

\title{Universal relations for fast rotating neutron stars without parametrisation bias}

\author[C. J. Kr\"uger, M. Celato]{
Christian J. Kr\"uger \orcid{0000-0003-2672-2055}$^{1,2}$\thanks{E-mail: chr.krueger@uni-tuebingen.de},
Mariachiara Celato \orcid{0009-0003-7478-2922}$^{1,3}$\thanks{E-mail: mariachiara.celato@uni-tuebingen.de}
\\
$^{1}$Theoretical Astrophysics, Institut f\"ur Astronomie und Astrophysik, University of T\"ubingen, 72076 T\"ubingen, Germany\\
$^{2}$High Performance and Cloud Computing Group ZDV, University of T\"ubingen, 72076 T\"ubingen, Germany\\
$^{3}$High-Energy Astrophysics, Institut f\"ur Astronomie und Astrophysik, University of T\"ubingen, 72076 T\"ubingen, Germany
}

\date{\today}

\begin{document}
\maketitle

\begin{abstract}
We provide a summary of several parametrisations for the nuclear equation of state that have been proposed over the past decades and list viable ranges for their parameters. Based on these parametrisations, we construct a large database of rotating neutron star models using the \textsc{rns} code, which are arranged in sequences of constant central energy density. After filtering these sequences with respect to generous astrophysical constraints, we discover tight universal relations for various bulk quantities of uniformly rotating neutron stars of arbitrary rotation rates. These universal relations allow bulk quantities of uniformly rotating neutron stars to be estimated at minimal computational cost, approximately three orders of magnitude faster than solving the full stellar equilibrium equations, employing only mass, radius, and moment of inertia of associated non-rotating neutron stars. Calibrated on a large, model-agnostic dataset, the relations are robust and remain accurate when tested against realistic equations of state and outperform previously published ones. For most quantities the 99th percentile of the relative residuals lies below 1.6\%; the moment of inertia (2.8\%) and T/W (4.6\%) carry the largest uncertainties. Such relations are important for future, high-precision measurements coming from electromagnetic and gravitational wave observations and may be used in equation of state inference codes or gravitational wave modeling among others.
\end{abstract}


\section{Introduction}

Neutron stars are natural laboratories for probing matter under extreme conditions that are unrivaled by terrestrial laboratories. Their structure is governed by the nuclear equation of state of dense matter---the microphysical input required to close the stellar-structure equations and, together with general relativity, fix an equilibrium configuration. The EoS relates pressure $p$, rest-mass density $\rho$, and specific entropy $\hat{s}$, typically written $p = p(\rho, \hat{s})$, and remains one of the central uncertainties in the field; constraining it through microphysical calculation and astronomical observation is a major goal of current astrophysics (for reviews, see, e.g., \citet{2016ARA&A..54..401O, 2021PrPNP.12003879B, 2021ARNPS..71..433L}).

Observational constraints have advanced rapidly. Pulsar timing of $2 M_\odot$ neutron stars \citep{2010Natur.467.1081D, 2013Sci...340..448A, 2020NatAs...4...72C, 2021ApJ...915L..12F} sets a strong lower bound on the maximum mass, while NICER pulse-profile modeling has yielded joint mass–radius constraints for PSR J0030+0451 and PSR J0740+6620 \citep{2019ApJ...887L..21R, 2021ApJ...918L..27R, 2019ApJ...887L..24M, 2021ApJ...918L..28M}. The binary neutron star merger GW170817 \citep{LIGOScientific:2017vwq, LIGOScientific:2017ync} constrained the tidal deformability and radius near $1.4 M_\odot$ \citep{2018PhRvL.120q2703A, 2018PhRvL.120z1103M, LIGOScientific:2018cki} and inaugurated multi-messenger studies of dense matter; next-generation detectors such as the Einstein Telescope \citep{2020JCAP...03..050M} and Cosmic Explorer \citep{2021arXiv210909882E} will sharpen these constraints further and heighten the need for accurate neutron star modeling.

Many equation-of-state models have been developed, ranging from phenomenological liquid-drop \citep{1971NuPhA.175..225B} and relativistic mean-field \citep{1974AnPhy..83..491W} descriptions to microscopic many-body calculations (e.g. \cite{1997A&A...328..274B, 1998PhRvC..58.1804A}) to high-density extensions including quark matter (e.g. \cite{1974PhRvD...9.3471C}); for a comprehensive overview of the theoretical modeling of the equation of state across density regimes and its confrontation with observations, see \cite{2025RvMP...97d5007C}. Alongside these physically motivated models, several parametrised representations — the well-known piecewise polytropes \citep{2009PhRvD..79l4032R}, sound-speed–based approaches \citep{2019MNRAS.485.5363G, 2020NatPh..16..907A}, and Taylor expansions around saturation density \citep{2018PhRvC..97b5805M} — capture the essential features of an equation of state with only a few parameters, which may be fitted to a specific equation of state or sampled from priors in model-agnostic studies. We adopt the latter approach, presenting and comparing nine parametrisations drawn from the literature \citep{2009PhRvD..79l4032R, 2020PhRvD.102h3027O, 2010PhRvD..82j3011L, 2018PhRvD..97l3019L, 2019PhRvD.100j4048F, 2019MNRAS.485.5363G, 2020NatPh..16..907A}. Since each parametrisation, restricted to its literature-suggested number of parameters, introduces some bias, fitting data points from a wide range of parametrisations simultaneously mitigates such biases and yields robust, model-agnostic \emph{universal relations}.

Universal relations relate certain bulk quantities (or combinations thereof) of neutron stars in a way that is largely independent of the equation of state. Early examples include the relation between the normalised moment of inertia $I/MR^2$ and the compactness $C=M/R$ \citep{1994ApJ...424..846R}, later refined by \cite{2001ApJ...550..426L} and extended to rapidly rotating neutron stars by \cite{2016MNRAS.459..646B}. Also in the 1990s, \cite{1996PhRvL..77.4134A, 1998MNRAS.299.1059A} discovered the first universal relations concerning mode properties, opening up the field of gravitational-wave asteroseismology (revisited by \cite{1999MNRAS.310..797B, 2004PhRvD..70l4015B, 2005MNRAS.357.1029T}). Another example are the famous $I$-Love-$Q$ relations discovered by \cite{2013Sci...341..365Y}.

The majority of the universal relations have been concerned with non-rotating compact objects, both because of the computational cost of constructing rapidly rotating models and because of the enlarged parameter space. Since the late 2000s, advances in the seismology of rotating neutron stars \citep{2008PhRvD..78f4063G, 2009PhRvD..80f4026G, 2013PhRvD..88d4052D, 2020PhRvL.125k1106K} have produced an increasing number of such relations, including extensions of $I$-Love-$Q$ to rotating stars \citep{2014PhRvL.112l1101P, 2014PhRvL.112t1102C}, relations for the general shape of rotating stars~\citep{2014ApJ...791...78A, 2021PhRvD.103f3038S, 2023ApJ...954...16G, 2024ApJ...962...61M}, and relations along the Keplerian sequence \citep{2019PhRvD..99d3004R}. Data science-driven approaches have been reported recently as well \citep{2023PhRvD.107j3050P, 2024PhRvD.109j3033M}, as have relations concerning masses and radii of rotating neutron stars \citep{2022ApJ...934..139K}. Universal relations have also been established for hot, rapidly rotating merger remnants, where universality is recovered at fixed entropy per baryon and electron fraction and has been used to bound the maximum mass from GW170817 \citep{2020MNRAS.499..914R, 2021PhRvC.103e5811K}.

Beyond identifying the relevant aspects of a problem, universal relations are valuable for applications requiring heavy computation, such as inference of neutron star properties (e.g. \cite{2019PhRvD..99l3026K, 2021PhRvD.103f3036G}) or statistical parameter estimation over equation-of-state parameters (e.g. \cite{2022PhRvD.105l4071V, 2023PhRvD.108l4056K}). While mass and radius may be cheaply obtained for a given equation of state, more involved observables such as properties of $f$-mode oscillations cannot easily be computed during parameter estimation, which is where universal relations become indispensable.

This paper builds on our earlier work~\citep{2023PhRvD.108l4056K}, where we proposed accurate universal relations for the moment of inertia $I$ and the ratio of kinetic to gravitational binding energy $T/W$ of rotating neutron stars, based on the mass $M_\star$, radius $R_\star$ and moment of inertia $I_\star$ of the zero-spin star with the same central energy density $\epsilon_c$. Here, we overcome previous limitations by using a broader, model-agnostic equation-of-state dataset based on nine parametrisations constrained only by generous astrophysical considerations. This leads to slightly larger uncertainties but still yields robust and accurate relations. Since the non-rotating reference stars are easy to compute via ODE integration, and the relations themselves are simple analytical expressions, the resulting estimates for rapidly rotating stars are computationally cheap yet accurate to within a few percent.

The paper is organised as follows. In Sec.~\ref{sec:eos_params}, we discuss nine equation-of-state parametrisations which have been proposed in the literature and we provide prior ranges for their parameters. In Sec.~\ref{sec:crustal_eos}, we discuss different models for the crustal equation of state and present our choice that we use in this study. Section~\ref{sec:seq_const_eps_c} is devoted to sequences of constant central energy density, which are crucial for the development of our universal relations. In Sec.~\ref{sec:dataset}, we describe in detail how we build our database comprising a large number of rotating neutron star models across all presented equation-of-state parametrisations. Furthermore, we introduce the astrophysical constraints that we apply to our database in order to arrive at astrophysically relevant universal relations. In Sec.~\ref{sec:urs}, we present the discovered universal relation for various bulk quantities of rapidly and uniformly rotating neutron stars. These relations provide estimates with percent-level accuracy of a rotating neutron star employing mass, radius, and moment of inertia of an associated, non-rotating neutron star. Section~\ref{sec:comparison} presents comparisons of our relations to previously published ones and in Sec.~\ref{sec:ur-test} we perform tests to demonstrate their accuracy. We conclude the paper in Sec.~\ref{sec:summary} with a summary.

Unless noted otherwise, we employ units in which $c = G = M_\odot = 1$ throughout this paper.

\section{Equation-of-State Parametrisations}
\label{sec:eos_params}

The equation of state is a crucial ingredient in constructing equilibrium configurations of neutron stars, which has to be manually supplied. It encodes the thermodynamic properties of dense nuclear matter at the extreme densities encountered in the cores of neutron stars. In general, the equation of state is a relation between the pressure~$p$, rest-mass density~$\rho$, and specific entropy~$\hat{s}$, and is often expressed as a function $p = p(\rho, \hat{s})$; note that this particular choice of variables is not unique; the equation of state may be equivalently formulated using other thermodynamic quantities.

For most applications, a neutron star may be treated as cold, allowing thermal effects to be neglected. In this case, the equation of state becomes independent of the entropy and reduces to a one-parameter relation, typically written as $p = p(\rho)$; we will investigate only such equations of state here. While the equation of state is reasonably well constrained at low densities, $\rho \lesssim 2.7 \times 10^{14}\,\textrm{g}/\textrm{cm}^3$ (i.e., up to around nuclear saturation density) \citep{2010PhRvL.105p1102H}, it suffers from large uncertainties at higher densities. In fact, the pressure at a given density can vary by up to an order of magnitude across different models \citep{2009PhRvD..79l4032R}. In addition to many different physically motivated models for the equation of state, several parametrised representations have been developed to capture the essential features of an equation of state with only a few parameters. These parameters may be fitted to reproduce a specific equation of state, or, conversely, sampled from broad physical priors in model-agnostic studies.

The latter is what the present study is based on; we will discuss nine different equation-of-state parametrisations which have been developed with different applications in mind. Hence, they are based on approximating different thermodynamic variables in different ways, e.g., via an expansion ansatz, piecewise linear functions or similar; in addition to the pressure $p$ and the rest-mass density $\rho$, these parametrisations are also built on other thermodynamic variables. The energy density $\epsilon$ can be related to the former two variables by means of the first law of thermodynamics
\begin{equation}
    \dif \epsilon
    = \frac{\epsilon + p}{\rho} \dif \rho,
\end{equation}
where the prefactor of $\dif\rho$ on the right-hand side is also known as the specific enthalpy $h$, given by
\begin{equation}
    h = \frac{\epsilon + p}{\rho}.
    \label{eq:def_h}
\end{equation}
In some applications, it is convenient to introduce a pseudo-enthalpy $H$, defined by
\begin{equation}
    H(p) := \int_0^p \frac{\dif p'}{\epsilon(p') + p'}.
    \label{eq:H_expr1}
\end{equation}
Given this definition, it is easy to show that
\begin{equation}
    \dif H = \frac{1}{\epsilon + p} \dif p,
    \label{eq:H_expr2}
\end{equation}
and we also have $H = \ln(h)$\footnote{In fact, the two expressions Eq.~\eqref{eq:def_h} and \eqref{eq:H_expr1} for the pseudo-enthalpy and enthalpy do not result directly in $H = \ln(h)$ but instead in $H = \ln(h/h_0)$ with an integration constant $h_0$. However, in most applications, this constant is assumed to be $h_0 = 1$ \citep{2024arXiv240411346K}.}. Furthermore, we have the adiabatic index
\begin{equation}
    \gamma
    := \frac{\epsilon + p}{p} \left( \frac{\partial p}{\partial \epsilon} \right)_{\hat{s}}
    \label{eq:def_adiabatic_index}
\end{equation}
and the speed of sound
\begin{equation}
    c_s^2
    := \left( \frac{\partial p}{\partial \epsilon} \right)_{\hat{s}}
    = \frac{\gamma p}{\epsilon + p}.
\end{equation}

Usually, a particular equation-of-state parametrisation is given by expressing one of the thermodynamic variables as a power series, as a step function or approximating it in some other way; subsequently, the above definitions and relations are employed in order to reconstruct the remaining quantities.

In the following, we will briefly discuss the different parametrisations that we consider in this study. Each of them comes with a different, individual set of parameters. Note that different parametrisations may use the same variable albeit with different meanings. To stay consistent with the original publications, we keep the variable names unchanged; each subsection should be considered self-contained. For each parametrisation, we will show prior ranges for its parameters (cf.~Tabs.~\ref{tab:param-ppa} - \ref{tab:param_sos_annala}) from which we uniformly draw values for potential equations of state; additionally, we will also show ``viable ranges'' that lead to physically consistent models under our filtering procedure (see Sec.~\ref{sec:dataset}). These are shown here to provide immediate context, but they can be considered results of the astrophysical constraints.

In some cases, the quoted viable ranges are identical to the prior ranges, which may indicate that the prior range could have been larger. In certain cases, the prior ranges are physically motivated, giving them a solid foundation. For certain parametrisations (e.g., the spectral representations in Sec.~\ref{ssec:spectral}), we gradually extended the prior ranges until only isolated points appeared at the boundary, but refrained from further extension because of the computational effort required.

For each equation-of-state parametrisation, we illustrate its typical behaviour by showing representative mass--radius diagrams in App.~\ref{app:m-r}. All presented equation-of-state parametrisations have been implemented in a C library, using numerical routines from the GNU Scientific Library (GSL) \citep{gsl}.

\subsection{Piecewise Polytropes}
\label{sec:piecewise_polytropes}

A polytropic equation of state
\begin{align}
    p(\rho)
    & = K \rho^\Gamma,
\end{align}
where $K$ is a length scale, assumes a constant adiabatic index $\Gamma$ throughout the star, whereas $\Gamma$ is known to vary considerably between its layers \citep{1971ApJ...170..299B}. While the idea of modeling the equation of state in several pieces predates their work \citep{1999ApJ...525L..45H, 2000ARNPS..50..481H, 2001ApJ...550..426L}, \cite{2009PhRvD..79l4032R} introduced a systematic \emph{piecewise polytropic} parametrization, in which the adiabatic index is held constant within each of several density intervals delimited by fixed transition densities; in other words, $\Gamma(\rho)$ is considered to be a step function and the locations of the discontinuities are the ``dividing densities'' $(\rho_i)_{i=0}^n$, splitting the entire domain of densities into $n$ intervals. These dividing densities can be specified freely. In the $i$-th density range, the equation of state then has to obey
\begin{align}
    p(\rho)
    & = K_i \rho^{\Gamma_i}
    \quad\text{for}\quad
    \rho_{i-1} \le \rho < \rho_i.
\end{align}
Requiring the pressure $p$ and the energy density $\epsilon$ to be continuous across each dividing density fixes the segment constants $K_i$ and the integration constants $a_i$, yielding
\begin{align}
    \epsilon(\rho)
    & = (1 + a_i) \rho + \frac{K_i}{\Gamma_i - 1} \rho^{\Gamma_i}
    \quad\text{for}\quad
    \rho_{i-1} \le \rho < \rho_i.
\end{align}
It turns out that $\gamma = \Gamma_i$ within each interval. The parameters thus carry a direct physical meaning, and nearly all resulting expressions are analytical, which keeps the numerical implementation simple and computationally cheap.

\cite{2009PhRvD..79l4032R} showed that three polytropic segments covering the core reproduce the bulk properties of realistic equations of state (such as the maximum mass, the maximum rotation rate, and the radius $R_{1.4}$ of a $1.4\,M_\odot$ star) to within a few percent, specified by the four parameters $(p_1, \Gamma_1, \Gamma_2, \Gamma_3)$, where $p_1$ is the pressure at the first dividing density; this is the widely used piecewise polytropic approximation (PPA). We adopt uniform priors for these four parameters (Tab.~\ref{tab:param-ppa}), as also used by \cite{2018PhRvD..98f3004C}, with the two core dividing densities fixed to $10^{14.7}\,\gperccc$ and $10^{15}\,\gperccc$ as originally proposed by \cite{2009PhRvD..79l4032R}.

\begin{table}
    \centering
    \caption{Uniform priors for the parameters of the piecewise polytropic parametrisation (referred to as PPA), which have also been used by~\protect\cite{2018PhRvD..98f3004C}. The ``viable range'' indicates the subset of the prior space that leads to physically consistent equation of state models under our constraints (cf.~Sec.~\ref{sec:dataset}).}
    \begin{tabular}{lcc}
    \toprule
    Parameter & Prior Range & Viable Range \\
    \midrule
    $\log(p_1 / \dynpercc) $ & $[33.6,\, 35.4]$ & $[34.2, 34.8]$ \\
    $\Gamma_1$ & $[2.0,\, 4.5]$ & $[2.0, 4.5]$ \\
    $\Gamma_2$ & $[1.1,\, 4.5]$ & $[1.9, 4.1]$ \\
    $\Gamma_3$ & $[1.1,\, 4.5]$ & $[1.1, 2.9]$ \\
    \bottomrule
    \end{tabular}
    \label{tab:param-ppa}
\end{table}

\begin{figure}
    \centering
    \includegraphics[width=\linewidth]{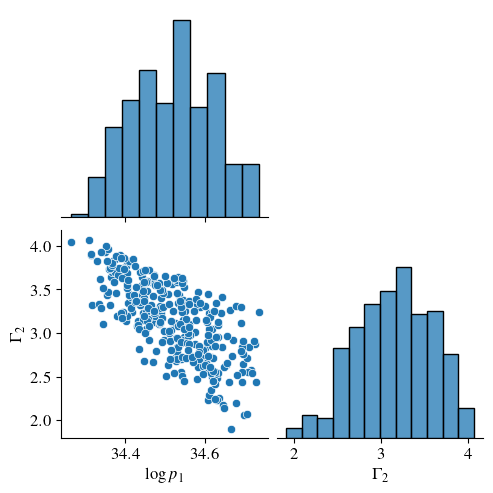}
    \caption{Scatter plot for the parameters $p_1$ and $\Gamma_2$ in the piecewise polytropic parametrization; the correlation coefficient is $-0.715$.}
    \label{fig:corr-ppa}
\end{figure}

The four PPA parameters are not independent once astrophysical constraints are imposed. Among the equations of state surviving our filtering (cf.~Sec.~\ref{sec:dataset}), $p_1$ and $\Gamma_2$ are anticorrelated, with a correlation coefficient of $-0.715$ (Fig.~\ref{fig:corr-ppa}). This is readily understood: since $p_1 = p(\rho_1)$ fixes the pressure at the first dividing density and the pressure at the second follows as $p(\rho_2) = p_1 (\rho_2/\rho_1)^{\Gamma_2}$, a higher $p_1$ must be compensated by a lower $\Gamma_2$ for $p(\rho_2)$ to remain astrophysically viable. This is worth bearing in mind when sampling the PPA parameters, as independent priors on $(p_1, \Gamma_2)$ may overpopulate physically inconsistent regions of the parameter space.

\subsection{Generalised Piecewise Polytropes}
\label{sec:gppa}

Piecewise polytropes do not enforce differentiability of the pressure or energy density at the dividing densities, so the sound speed will in general be discontinuous there, which may introduce spurious artifacts in hydrodynamics simulations. The generalised piecewise polytropic approximation (GPPA) proposed by \cite{2020PhRvD.102h3027O} removes this by adding a constant offset $\Lambda_i$ to each segment,
\begin{align}
    p(\rho)
    & = K_i \rho^{\Gamma_i} + \Lambda_i,
\intertext{with a correspondingly shifted energy density}
    \epsilon(\rho)
    & = (1 + a_i) \rho + \frac{K_i}{\Gamma_i - 1} \rho^{\Gamma_i} - \Lambda_i,
\end{align}
for $\rho_i \le \rho < \rho_{i+1}$. The extra constant $\Lambda_i$ is fixed by additionally requiring the pressure (and hence the energy density) to be differentiable at each dividing density, resulting in a continuous sound speed.

Following \cite{2009PhRvD..79l4032R}, this parametrisation uses three core segments, too, each with its own index $\Gamma_i$. However, the high-density equation of state is now shifted through $K_1$ rather than $p_1$; a GPPA is thus specified by $(\log K_1, \Gamma_1, \Gamma_2, \Gamma_3)$. Whereas \cite{2020PhRvD.102h3027O} originally matches a five-segment SLy crust \citep{2001A&A...380..151D}, we use the same single-segment crust ($\Gamma = 1.35692$; cf.~Sec.~\ref{sec:crustal_eos}) as for the other parametrisations, so as not to bias the comparison. We assume uniform priors for the four parameters (Tab.~\ref{tab:param_gppa}), with core dividing densities $10^{14.87}\,\gperccc$ and $10^{14.99}\,\gperccc$ as originally proposed \citep{2020PhRvD.102h3027O}.

\begin{table}
    \centering
    \caption{Uniform priors for the parameters of the generalised piecewise polytropes (GPPA).}
    \begin{tabular}{lcc}
    \toprule
    Parameter & Prior Range & Viable Range \\
    \midrule
    $\log(K_1)$ & $[-45, -20]$ & $[-45, -20]$\\
    $\Gamma_1$ & $[2.2, 4.0]$ & $[2.2, 4.0]$ \\
    $\Gamma_2$ & $[1.8, 4.0]$ & $[1.8, 4.0]$ \\
    $\Gamma_3$ & $[1.1, 4.0]$ & $[1.1, 3.1]$ \\
    \bottomrule
    \end{tabular}
    \label{tab:param_gppa}
\end{table}

\begin{figure}
    \centering
    \includegraphics[width=\linewidth]{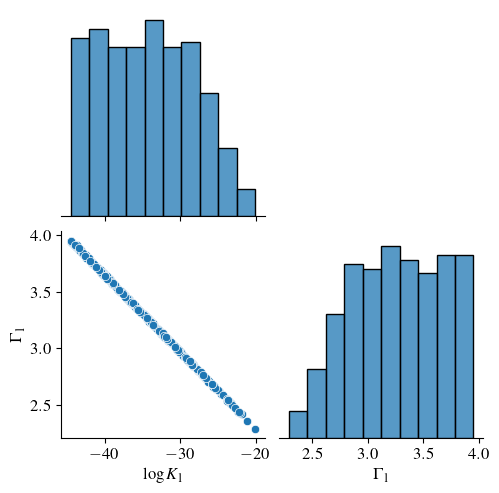}
    \caption{Scatter plot for the parameters $K_1$ and $\Gamma_1$ of 320 randomly drawn equations of state in the generalised piecewise polytropic parametrization (after filtering for astrophysically relevant equations of state); the correlation coefficient is $-0.999916$.}
    \label{fig:corr-gppa}
\end{figure}

After filtering for astrophysically relevant equations of state, we observe a near-perfect anticorrelation between $K_1$ and $\Gamma_1$, with a correlation coefficient of $-0.999916$ (Fig.~\ref{fig:corr-gppa}); the two nominally independent parameters are therefore nearly degenerate. This originates in the crust--core matching: $K_1$ and $\Gamma_1$ jointly determine the transition density
\begin{align}
    \rho_{\rm cc}
    & = \left( \frac{K_1 \Gamma_1}{K_{\rm crust} \Gamma_{\rm crust}} \right)^{1/ (\Gamma_{\rm crust} - \Gamma_1)},
\end{align}
where $K_{\rm crust}$ and $\Gamma_{\rm crust}$ are the parameters of the crustal segment adjacent to the transition. The density $\rho_{\rm cc}$ must lie between the crust's outermost dividing density (in the approximation to the crustal SLy equation of state) and the first core dividing density, i.e., $\rho_{\rm cc} \in [10^{11.73}, 10^{14.87}]\,\gperccc$. Equivalently, the value of $\log K_1$ is confined to a narrow, $\Gamma_1$-dependent band; defining $\Delta_{K_1} := \log_{10} K_{\rm 1,max} - \log_{10} K_{\rm 1,min}$, we have
\begin{align}
    \Delta_{K_1}
    & = (\Gamma_1 - \Gamma_{\rm crust})\, \log_{10}\left( \rho_{\rm cc,max} / \rho_{\rm cc,min}\right)
    \\
    & \le 3.14 \times (\Gamma_1 - \Gamma_{\rm crust}),
    \label{eq:Delta_K1}
\end{align}
with $\Gamma_{\rm crust} = 1.35692$. Additionally, the lower bound of $10^{11.73}\gperccc$ for $\rho_{\rm cc}$ is purely mathematically motivated and low values of $\rho_{\rm cc}$ often cause the speed of sound to become superluminal at relatively low densities, or lead to equations of state that produce unrealistically massive neutron stars. In practice, therefore, $\rho_{\rm cc,min}$ should lie considerably closer to its upper limit $\rho_{\rm cc,max}$, further narrowing the viable range of $K_1$ and accounting for the near-perfect correlation observed here.

\subsection{Spectral Representation of the High-Density Equation of State}
\label{ssec:spectral}

\subsubsection{Pressure-Based Representation for the Adiabatic Index}
\label{sssec:pb_s}

Shortly after the piecewise polytropes, \cite{2010PhRvD..82j3011L} proposed a parametrisation in which, instead of a piecewise-constant adiabatic index that jumps at the segment boundaries, the index is a smooth function $\gamma(p)$ of the pressure above the crust--core matching pressure $p_0$. For a given $\gamma(p)$, integrating the definition of the adiabatic index, cf.~Eq.~\eqref{eq:def_adiabatic_index}, reconstructs the equation of state as \citep{2010PhRvD..82j3011L}
\begin{align}
    \epsilon(p)
    & = \frac{\epsilon_0}{\mu(p)} + \frac{1}{\mu(p)} \int_{p_0}^p \frac{\mu(p')}{\gamma(p')} \dif p',
    \label{eq:pb_eps_p}
\intertext{where $\mu(p)$ is an auxiliary function given by}
    \mu(p)
    & = \exp \left[ - \int_{p_0}^p \frac{\dif p'}{p' \gamma(p')} \right].
    \label{eq:pb_mu_p}
\end{align}
with the integration constant $\epsilon_0$ fixed by continuity at the matching point, $\epsilon_0 = \epsilon(p_0)$. Step-function choices for $\gamma(p)$ recover the piecewise polytropes of Sec.~\ref{sec:piecewise_polytropes} as a special case; Lindblom instead expands $\gamma(p)$ spectrally for $p \ge p_0$,
\begin{align}
    \gamma(p)
    & = \exp \left( \sum_{k=0}^\infty \gamma_k \Phi_k(p) \right)
    \quad\text{with}\quad
    \Phi_k(p) = \left[ \ln (p/p_0) \right]^k,
    \label{eq:pb_s_gamma}
\end{align}
with spectral coefficients $\gamma_k$. We refer to this pressure-based representation as PB-S \citep{2018PhRvD..97l3019L} and list the uniform prior ranges for the $\gamma_k$ in Tab.~\ref{tab:param_pb_s}.

\begin{table}
  \centering
  \caption{Ranges for the parameters $\gamma_k$ for the pressure-based expansion of $\gamma(p)$ (referred to as PB-S) as given in Eq.~\eqref{eq:pb_s_gamma}.}
  \begin{tabular}{ccc}
  \toprule
  Parameter & Prior Range & Viable Range \\
  \midrule
  $\gamma_0$ & $[\hphantom{-}0.20, 2.00]$ & $[\hphantom{-}0.20, 2.00]$ \\
  $\gamma_1$ & $[-1.60, 1.70]$ & $[-0.80, 1.65]$ \\
  $\gamma_2$ & $[-0.60, 0.60]$ & $[-0.50, 0.25]$ \\
  $\gamma_3$ & $[-0.02, 0.04]$ & $[-0.02, 0.04]$ \\
  \bottomrule
  \end{tabular}
  \label{tab:param_pb_s}
\end{table}

Because the basis functions are the non-orthogonal monomials $\left[\ln(p/p_0)\right]^k$, the coefficients $\gamma_k$ are strongly intercorrelated (a larger $\gamma_0$ is compensated by a smaller $\gamma_1$, and the coefficients decrease in magnitude with order as higher-order terms act more strongly at large argument). These correlations are of purely mathematical origin, so we do not assign them physical meaning or show scatter plots; they are also immaterial to the representation's power, since keeping arbitrarily many terms allows any smooth equation of state to be reproduced. They do, however, have a practical consequence: sampling the $\gamma_k$ from independent uniform priors yields a high rejection rate under our astrophysical constraints, with only about $0.5$--$1.0\%$ of drawn quadruples surviving (for both the pressure- and enthalpy-based expansions). The same remarks apply to the enthalpy-based variant below, which shares the same expansion structure.

\subsubsection{Enthalpy-Based Representation for the Adiabatic Index}

The same construction can be carried out in the pseudo-enthalpy $H$ rather than the pressure. This is convenient because the enthalpy-based form of the TOV equations \citep{1992ApJ...398..569L} requires $\epsilon$ and $p$ as functions of $H$, and because the \textsc{rns} code we use takes the pseudo-enthalpy as its central thermodynamic variable \citep{1995ApJ...444..306S, rnscode}. For a given $\gamma(H)$, the equation of state follows from
\begin{align}
    \dif p
    & = (\epsilon + p) \dif H,
    \\
    \dif \epsilon
    & = \frac{(\epsilon + p)^2}{p \gamma(H)} \dif H,
\end{align}
whose general solution is given in Eqs.~(17)--(19) by \cite{2010PhRvD..82j3011L}. As before, we expand the index spectrally,
\begin{align}
    \gamma(H)
    & = \exp \left( \sum_{k=0}^\infty \gamma_k \Phi_k(H) \right)
    \quad\text{with}\quad
    \Phi_k(H) = \left[ \ln (H/H_0) \right]^k,
    \label{eq:hb_s_gamma}
\end{align}
with $H_0 = H(p_0)$ the pseudo-enthalpy at the crust--core matching point. We refer to this variant as HB-S \citep{2018PhRvD..97l3019L}; the uniform prior ranges for the $\gamma_k$ are listed in Tab.~\ref{tab:param_hb_s}.

Both spectral representations yield smooth core equations of state with rapidly converging series, and \cite{2010PhRvD..82j3011L} showed that the physical adiabatic index is approximated more efficiently by such a decomposition than by a step function. Their coefficients $\gamma_k$ lack the intuitive meaning of the $\Gamma_i$ of the piecewise polytropes, the sole exception being $\gamma_0 = \gamma(H_0)$, the adiabatic index at the matching point.

\begin{table}
  \centering
  \caption{Ranges for the parameters $\gamma_k$ for the enthalpy-based expansion of $\gamma(H)$ (referred to as HB-S) as given in Eq.~\eqref{eq:hb_s_gamma}.}
  \begin{tabular}{ccc}
  \toprule
  Parameter & Prior Range & Viable Range \\
  \midrule
  $\gamma_0$ & $[\hphantom{-}0.20, 2.00]$ & $[\hphantom{-}0.40, 2.00]$ \\
  $\gamma_1$ & $[-2.00, 2.80]$ & $[-1.60, 2.00]$ \\
  $\gamma_2$ & $[-0.80, 1.00]$ & $[-0.80, 1.00]$ \\
  $\gamma_3$ & $[-0.25, 0.25]$ & $[-0.25, 0.15]$ \\
  \bottomrule
  \end{tabular}
  \label{tab:param_hb_s}
\end{table}

\subsubsection{Pressure-Based Speed-of-Sound Representation}
\label{ssec:spectral_cs2_p}

The parametrisations above expand the adiabatic index and thereby guarantee only $\gamma > 0$; they place no bound on the sound speed, so a given choice of coefficients may yield an equation of state that becomes acausal at densities relevant for neutron star matter. To enforce causality by construction, \cite{2018PhRvD..97l3019L} expands instead the \emph{velocity function}
\begin{align}
    \Upsilon
    & = \frac{1 - c_s^2}{c_s^2},
\end{align}
since any non-negative $\Upsilon$ corresponds to a physical sound speed $0 \le c_s^2 < 1$. For $\Upsilon = \Upsilon(p)$, the equation of state follows from $\dif\epsilon = [1 + \Upsilon(p)]\,\dif p$, i.e.
\begin{align}
    \epsilon(p)
    & = \epsilon_0 + p - p_0 + \int_{p_0}^p \Upsilon(p')\, \dif p',
\end{align}
where $\epsilon_0$ is the integration constant given by $\epsilon_0 = \epsilon(p_0)$. As before, $\Upsilon(p)$ is expanded in the same non-orthogonal monomial basis (cf.~Sec.~\ref{sssec:pb_s}),
\begin{align}
    \Upsilon(p)
    & = \exp \left( \sum_{k=0}^\infty \upsilon_k \Phi_k(p) \right)
    \quad\text{with}\quad
    \Phi_k(p) = \left[ \ln (p/p_0) \right]^k,
    \label{eq:pb_c_s_gamma}
\end{align}
which we refer to as PB-C-S \citep{2018PhRvD..97l3019L}. The uniform prior ranges for the $\upsilon_k$ are given in Tab.~\ref{tab:param_pb_c_s}.

\begin{table}
    \centering
    \caption{Ranges for the parameters $\upsilon_k$ for the pressure-based expansion of $\Upsilon(p)$ (referred to as PB-C-S) as given in Eq.~\eqref{eq:pb_c_s_gamma}.}
    \begin{tabular}{lcc}
    \toprule
    Parameter & Prior Range & Viable Range \\
    \midrule
    $\upsilon_0$ & $[-10.0, 7.0]$ & $[-10.0, 7.0]$ \\
    $\upsilon_1$ & $[-12.0, 15.0]$ & $[-10.8, 14.9]$ \\
    $\upsilon_2$ & $[-8.0, 8.0]$ & $[-6.7, 4.5]$ \\
    $\upsilon_3$ & $[-1.5, 1.0]$ & $[-0.6, 0.7]$ \\
    \bottomrule
    \end{tabular}
    \label{tab:param_pb_c_s}
\end{table}

\subsubsection{Enthalpy-Based Speed-of-Sound Representation}
\label{ssec:spectral_cs2_H}

The velocity function may equally be expanded in the pseudo-enthalpy, $\Upsilon = \Upsilon(H)$, in which case the equation of state follows from
\begin{align}
    \dif \epsilon
    & = (\epsilon + p) (1 + \Upsilon(H)) \dif H,
    \\
    \dif p
    & = (\epsilon + p) \dif H,
\end{align}
with the general solution given in Eqs.~(23)--(25) in \cite{2018PhRvD..97l3019L}. Expanding $\Upsilon(H)$ in the same basis,
\begin{align}
    \Upsilon(H)
    & = \exp \left( \sum_{k=0}^\infty \upsilon_k \Phi_k(H) \right)
    \quad\text{with}\quad
    \Phi_k(H) = \left[ \ln (H/H_0) \right]^k,
    \label{eq:hb_c_s_gamma}
\end{align}
defines the variant we call HB-C-S; its uniform prior ranges are listed in Tab.~\ref{tab:param_hb_c_s}.

\begin{table}
    \centering
    \caption{Ranges for the parameters $\upsilon_k$ for the enthalpy-based expansion of $\Upsilon(H)$ (referred to as HB-C-S) as given in Eq.~\eqref{eq:hb_c_s_gamma}.}
    \begin{tabular}{lcc}
    \toprule
    Parameter & Prior Range & Viable Range \\
    \midrule
    $\upsilon_0$ & $[0.0, 6.0]$ & $[0.0,\ 6.0]$ \\
    $\upsilon_1$ & $[-15.0, 10.0]$ & $[-14.7, 8.9]$ \\
    $\upsilon_2$ & $[-10.0, 10.0]$ & $[-10.0, 10.0]$ \\
    $\upsilon_3$ & $[-3.0, 5.0]$ & $[-2.5, 3.7]$ \\
    \bottomrule
    \end{tabular}
    \label{tab:param_hb_c_s}
\end{table}

We note that these velocity-function expansions have recently been extended to Chebyshev-polynomial basis functions, for both the pressure- and enthalpy-based formulations \citep{2024PhRvD.110h3030L}. Since the Chebyshev polynomials form a complete set on $[-1, 1]$, they require the high-density domain to be rescaled to this interval and hence a maximum pressure $p_\mathrm{max}$ (or enthalpy $H_\mathrm{max}$) to be fixed in advance. Such a bound is natural when approximating a specific equation of state, but not in an agnostic setting like ours, so we do not include this parametrisation in our study.

\subsubsection{Continuity of the First Derivative}

The spectral representations above are matched to the crust at $p_0$ without any condition on the coefficients, so the derivatives of thermodynamic quantities---and thus the sound speed itself---may be discontinuous there. \cite{2022PhRvD.105f3031L} showed that fixing the leading coefficient,
\begin{align}
    \gamma_0 = \ln \Gamma_0
    \qquad \text{or} \qquad
    \upsilon_0 = \ln \left[ \frac{1 - c_{s,0}^2}{c_{s,0}^2} \right]
\end{align}
for the index- and velocity-function expansions respectively, where $\Gamma_0$ and $c_{s,0}$ are the adiabatic index and sound speed at $p_0$, renders the first derivatives (and hence the sound speed) continuous and improves the accuracy of the representations. The condition is identical for the pressure- and enthalpy-based forms.

\subsection{Smooth Equation-of-State Parametrisations for High-Accuracy Simulations}

Building on Lindblom's approach \citep{2010PhRvD..82j3011L}, \cite{2019PhRvD.100j4048F} (which we refer to as F19) again decompose the adiabatic index spectrally, but expand in the rest-mass density rather than the pressure or enthalpy, using the variable $x := \ln(\rho/\rho_0)$ with $\rho_0$ a reference density (typically the crust-core matching density). With $\gamma := \dif\ln p/\dif x$, the pressure and energy density follow as
\begin{align}
    p(x)
    & = p_0 \exp\!\left( \int_0^x \gamma(x')\, \dif x' \right),
    \label{eq:spec_smooth_p}
    \\
    \epsilon(x)
    & = \epsilon_0\, e^x + e^x \int_0^x p(\xi)\, e^{-\xi}\, \dif\xi,
\end{align}
with $p_0 := p(0)$ and $\epsilon_0 := \epsilon(0)$ at the matching point; further relations are given by \cite{2019PhRvD.100j4048F}. The index is expanded in monomials,
\begin{align}
    \gamma(x) & = \sum_{k=0}^N \gamma_k x^k,
    \label{eq:spec_smooth_gamma}
\end{align}
for which the integral in Eq.~\eqref{eq:spec_smooth_p} is analytic. In contrast to \cite{2010PhRvD..82j3011L}, the first two coefficients are fixed to $\gamma_0 = \Gamma_0$ and $\gamma_1 = 0$, with $\Gamma_0$ the crustal adiabatic index at the matching point, so that discontinuities are pushed to the third derivative of the pressure rather than the first. Truncating after the fourth term leaves the four parameters $(\rho_0, p_0, \gamma_2, \gamma_3)$; for our single-segment crust $\Gamma_0 = 1.35692$ (cf.~Sec.~\ref{sec:crustal_eos}).\footnote{\cite{2019PhRvD.100j4048F} allow $\Gamma_0$ to be a fifth free parameter; we instead fix $\Gamma_0 = 1.35692$ to match the polytropic crust used for all other parametrisations (cf.~Sec.~\ref{sec:crustal_eos}).} The uniform priors are given in Tab.~\ref{tab:param_spec_smooth}.

\begin{table}
    \centering
    \caption{Ranges for the parameters $p_0$, $\rho_0$, $\gamma_2$, and $\gamma_3$ for the expansion of $\gamma(x)$ as given in Eq.~\eqref{eq:spec_smooth_gamma} (the F19 parametrisation).}
    \begin{tabular}{lccc}
    \toprule
    Parameter & Prior Range & Viable Range & Unit \\
    \midrule
    $p_0$ & $\left[4.2,\, 30.0\right]$& $\left[4.2,\, 30.0\right]$ & $10^{31}$ \dynpercc \\
    $\rho_0$ & $\left[2.5,\, 8.2\right]$ & $\left[2.7,\, 7.3\right]$ & $10^{13}$ \gperccc \\
    $\gamma_2$ & $[0.7,\, 1.4]$ & $[0.7,\, 1.4]$ & \\
    $\gamma_3$ & $[-0.4,\, -0.1]$ & $[-0.4,\, -0.16]$ & \\
    \bottomrule
    \end{tabular}
    \label{tab:param_spec_smooth}
\end{table}

As for the other spectral expansions, $\gamma_2$ and $\gamma_3$ are fairly strongly correlated; this is again of purely mathematical origin (cf.~the last paragraph of Sec.~\ref{sssec:pb_s}).

\subsection{Speed-of-Sound parametrisation}
\label{sec:sos-greif}

For most of the parametrisations above it is easy to choose parameters that render the equation of state unphysical, e.g.\ superluminal. Lindblom's velocity-function expansion (Sec.~\ref{ssec:spectral}) was the first to enforce causality, $0 \le c_s^2 < 1$, by construction; the parametrisation of \cite{2019MNRAS.485.5363G} additionally builds in the asymptotic behaviour expected from perturbative QCD, namely that the sound speed approaches $c_s^2 \to 1/3$ from below at high density \citep{2010PhRvD..81j5021K, 2014ApJ...781L..25F}. Their six-parameter ansatz (which we call G19) reads
\begin{align}
    c_s^2(x)
    & = a_1 e^{-\frac{1}{2} \left( x - a_2 \right)^2/a_3^2}
        + a_6
        + \frac{\frac{1}{3} - a_6}{1 + e^{- a_5 (x - a_4)}},
    \label{eq:sos_greif_cs2_of_x}
\end{align}
where $x = \epsilon / (m_n n_0)$ is the normalised energy density, and $m_n$ is the baryon mass. The logistic term (third) approaches the pQCD limit $1/3$ from below, the Gaussian (first) models an intermediate-density rise of the sound speed, and $a_6$ is fixed by continuous matching to the crust near saturation density $n_0 = 0.16\,\textrm{fm}^{-3}$. The free parameters $a_1$ to $a_5$ shape the logistic and the Gaussian. The equation of state follows by integrating $\dif p = c_s^2(\epsilon)\, \dif\epsilon$ from the crust--core matching point.

For some choices of $a_1$ to $a_5$, it may happen that $c_s^2(x)$ as per Eq.~\eqref{eq:sos_greif_cs2_of_x} turns negative within a single interval $(x_0, x_1)$, which will be located to the right of the Gaussian ($x_0 > a_2$). In this case, the speed of sound will then be set to 0 within this interval, which physically corresponds to a density discontinuity. In the present study, we exclude such equations of state from our dataset.

\begin{table}
    \centering
    \caption{Ranges for the parameters $a_i$ of the speed-of-sound parametrisation G19 as given in Eq.~\eqref{eq:sos_greif_cs2_of_x}. These intervals were proposed by~\protect\cite{2022PhRvD.105d3016L}.}
    \begin{tabular}{lcc}
    \toprule
    Parameter & Prior Range & Viable Range \\
    \midrule
    $a_1$ & $[0.5, 1.5]$ & $[0.5, 1.3]$ \\
    $a_2$ & $[1.3, 5.0]$ & $[1.3, 5.0]$ \\
    $a_3$ & $[0.05, 3.0]$ & $[0.25, 3.0]$ \\
    $a_4$ & $[1.5, 21.0]$ & $[1.6, 21.0]$ \\
    $a_5$ & $[0.1, 1.0]$ & $[0.1, 1.0]$ \\
    \bottomrule
    \end{tabular}
    \label{tab:param_sos_greif}
\end{table}

We assume uniform priors for the $a_i$ (Tab.~\ref{tab:param_sos_greif}, as proposed by \cite{2022PhRvD.105d3016L}). After filtering, the only relevant correlation is a mild positive one between $a_2$ and $a_3$, which are the Gaussian's centre and width. Since a wide Gaussian is admissible only when it sits far enough (via $a_2$) from the crust matching point. This single-Gaussian form respects both causality and the pQCD limit; while it cannot represent every equation of state, it qualitatively reproduces the expected sound-speed profile and is what we use here.

\subsection{Speed-of-Sound parametrisation as a function of chemical potential}
\label{sec:sos-annala}

To probe whether a quark phase is present in massive neutron star cores, \cite{2020NatPh..16..907A} introduced a further speed-of-sound parametrisation (which we refer to as A20) built on the chemical potential $\mu$. The sound speed is specified through $N$ nodes $\left((\mu_i, c_{s,i}^2)\right)_{i=1}^N$ with $\mu_1 < \cdots < \mu_N$ and linear interpolation of $c_s^2(\mu)$ in between; the number density and pressure then follow from \cite{2020NatPh..16..907A} as
\begin{align}
    n(\mu)
    & = n_m \exp\!\left[ \int_{\mu_1}^{\mu} \frac{\dif \mu'}{\mu'\, c_s^2(\mu')} \right],
    \quad
    p(\mu) = p_m + \int_{\mu_1}^{\mu} n(\mu')\, \dif \mu',
\end{align}
where $n_m$ and $p_m$ are the number density and pressure at the crust--core matching point $\mu_1$. A pQCD parameter $X_\textrm{pQCD} \in [1, 4]$ sets the high-density limits of the sound speed and number density \citep{2014ApJ...781L..25F}.

\begin{table}
    \centering
    \caption{Uniform priors for the parameters of the speed-of-sound parametrisation A20 as described in Sec.~\ref{sec:sos-annala}. After drawing random values for the pairs $(\mu_i, c_{s,i}^2)$, they are ordered by increasing chemical potential $\mu_i$. We do not show viable ranges for this parametrisation, as all parameters can be drawn from the prior ranges with no additional constraints; further constraints on the parameter ranges will stem from more detailed microphysical calculations rather than astrophysical bounds.}
    \begin{tabular}{lcc}
    \toprule
    Parameter & Prior Range & Unit \\
    \midrule
    $X_\textrm{pQCD}$ & $[1.0,\ 4.0]$ & \\
    $\mu_i$ & $[0.95,\ 2.60]$ & GeV \\
    $c_{s,i}^2$ & $[0.0,\ 1.0]$ & \\
    \bottomrule
    \end{tabular}
    \label{tab:param_sos_annala}
\end{table}

The two endpoints are fixed: $\mu_1$ is the crust--core matching potential\footnote{\cite{2020NatPh..16..907A} constrain the crust via chiral effective field theory up to $n_\textrm{CET} = 1.1\,n_0$ and set $\mu_1 = \mu(n_\textrm{CET})$; this density can be varied somewhat, and the parametrisation may equally be matched to a crustal equation of state at a slightly different $\mu_1$.} and $\mu_N = 2.6\,\textrm{GeV}$ ($\approx 40\,n_0$, where pQCD applies), with their sound speeds and pressures set by the crust and pQCD, respectively. Requiring the number density to be continuous at both matching points fixes two further parameters; we choose to constrain the two uppermost sound speeds $c_{s,N-2}^2$ and $c_{s,N-1}^2$. Hence, of the $N$ pairs plus $X_\textrm{pQCD}$, a total of $2N-5$ parameters are free; for this study, we chose $N=6$.

\section{Crustal Equation of State}
\label{sec:crustal_eos}

Our focus in this study lies on the high density nuclear equation of state, which suffers from large uncertainties and for which we have presented nine different parametrisations in the prior sections. The crustal equation of state, however, is fairly well constrained from terrestrial experiments up to about saturation density $n_0 = 0.16\,\textrm{fm}^{-3}$.

In the literature, there are several approaches as to how the crust is modelled, depending on the application and specific needs of the investigation. A more precise approach is to use tabulated data where all required quantities are listed to the desired precision; these data may include detailed information such as atomic numbers, density discontinuities due to abrupt changes in the chemical composition, shear moduli or other quantities. In between the data points, one needs to interpolate in a thermodynamically consistent way in order to construct a neutron star model. Such models are employed when the properties of the crust are under investigation, or when an accurate representation of the crustal composition is essential to the outcome of the study---for example, in modelling the composition of ejecta from a binary merger.

However, as consistent interpolation may not be straightforward to achieve and adds computational expense to the problem, crustal models are often simplified to reduce the overall complexity of the problem. The piecewise polytropic approximation (cf.~Sec.~\ref{sec:piecewise_polytropes}) does not only cater for the core of the neutron star but also provides a four-segment approximation to the SLy crustal equation of state proposed by \cite{2001A&A...380..151D}, which improves the accuracy over a simple polytropic crust but keeps most of the thermodynamical relations analytic and hence computationally cheap. Similarly, its extension to the generalised piecewise polytropic approximation (cf.~Sec.~\ref{sec:gppa}) also comes with suggested coefficients to approximate a realistic crust within its framework. The simplest way, however, to approximate the crustal equation of state is to simply use a polytropic equation of state, and this approach is still widely used, e.g., in binary merger simulations for gravitational wave forms, where computational expense and smoothness of the solution is critical but the impact of the crust is mostly negligible. This approach has explicitly been chosen in the F19 parametrisation where polytropic crusts with $\Gamma = 1.35692$ and $\Gamma = 2$ are employed \citep{2019PhRvD.100j4048F}.

When comparing different parametrisations or averaging over them, as we do in the present study, it is crucial to use the same crustal equation of state across all cases. To this end, we adopt a polytropic crust described by $p = K \rho^\Gamma$ with $\Gamma = 1.35692$. This adiabatic index is that of the highest density segment of the crust as proposed by \cite{2009PhRvD..79l4032R} and has been used in several studies (cf., e.g., \cite{2009PhRvD..79l4033R, 2010PhRvD..82d4049K, 2015PhRvD..91l4041D, 2019PhRvD.100j4048F, 2020PhRvD.101j3008C}). The polytropic constant $K$ is chosen such that the crust-core transition is continuous. This single-segment polytrope is currently implemented uniformly across all our high-density parametrisations; matching each to a tabulated crust in a thermodynamically consistent manner is a natural extension that we leave to future work. Since the crust mass is small compared to the total mass of the neutron star, its effect on the universal relations for bulk quantities is negligible---except for the equatorial radius $R_e$: absolute radii are known to be sensitive to the crustal equation of state and its matching to the core model \citep{2016PhRvC..94c5804F}, and relative to models with a more realistic crust our models can show deviations of up to a few hundred metres. However, since we report universal relations between bulk quantities rather than absolute radii, a deviation in $R_e$ that remains nearly constant along most of the stellar sequence largely cancels, so we expect the effect on the final universal relations to be small. The largest deviation due to a more realistic crust appears in the estimate of the Keplerian value $R_{e,K}$, since the radius changes most significantly near the mass-shedding limit. For the two unified-crust equations of state tested in Sec.~\ref{sec:ur-test}, APR4 and DDME2, the residuals of the radius relations remain below 2\% even there, comparable to the typical accuracy of our relations. While two cases cannot establish a general bound, they support our expectation that the crust treatment does not compromise the radius relations.

\section{Sequences of Constant Central Energy Density}
\label{sec:seq_const_eps_c}

\begin{figure}
    \centering
    \includegraphics[width=1\linewidth]{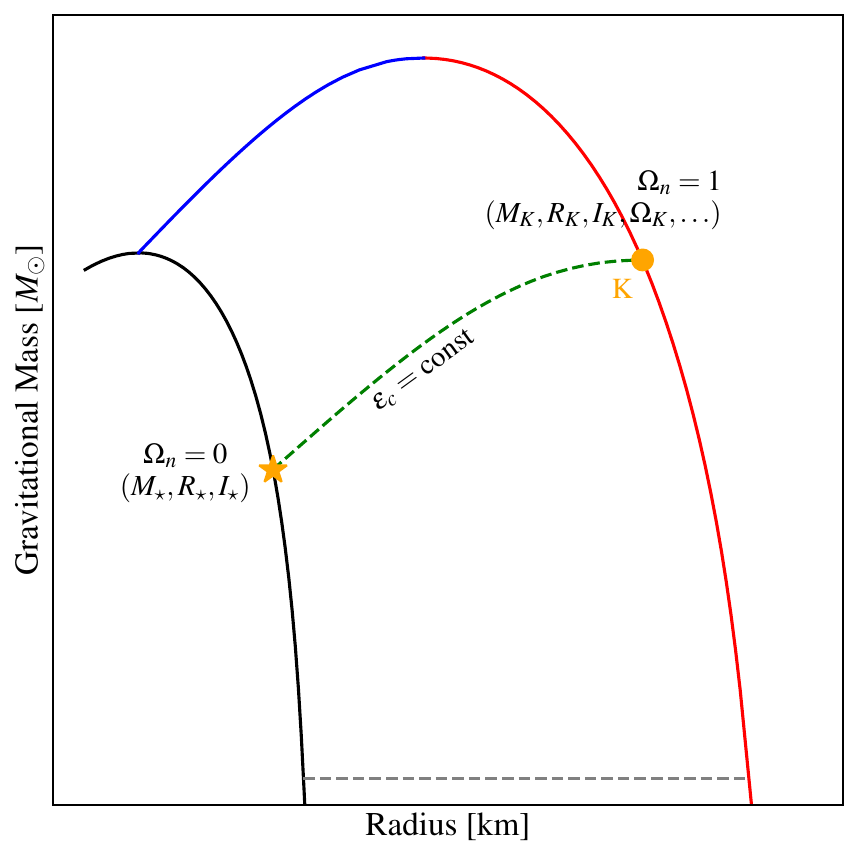}
    \caption{Visualisation of an exemplary sequence of constant central energy density (dashed, green) in a generic mass-radius diagram. Quantities belonging to the non-rotating model at $\Omega_n = 0$ are denoted with ``$\star$'', while Keplerian values at $\Omega_n = 1$ carry the subscript ``$K$''. The black line is the usual mass-radius curve of non-rotating neutron stars, the red line is the corresponding curve at the Kepler limit, the blue curve depicts the limit of quasi-radial instability, and the grey dashed line is our lower limit of $1.05\,M_\odot$. We do not show any labels on the axes as such diagrams are qualitatively identical for any equation of state.}
    \label{fig:sequence_def}
\end{figure}

As in our prior work \citep{2023PhRvD.108l4056K}, we focus on sequences of neutron stars along which the central energy density $\epsilon_c$ is held constant. The angular rotation rate $\Omega = 2\pi f_\textrm{spin}$ is varied and for a given $\epsilon_c$, it is bounded by the non-rotating limit ($\Omega = 0$) and the mass-shedding (or Kepler) limit, where the neutron star rotates at the maximum possible rotation rate $\Omega = \Omega_K$. Here and henceforth, we use the subscript ``K'' to denote a quantity that belongs to a model rotating at the Kepler limit; further, we use the subscript ``$\star$'' for quantities of the non-rotating member of the sequence. We show a visualisation of such a sequence for clarity in Fig.~\ref{fig:sequence_def}.

Such sequences are numerically easy to generate since the central energy density $\epsilon_c$ is typically one of the parameters that has to be specified in numerical codes which construct neutron star equilibrium configurations; in particular the \textsc{rns} code which we employ \citep{1995ApJ...444..306S, rnscode}. After specifying an equation of state, i.e., choosing one of the parametrisations discussed above and a set of corresponding parameters, we employ our modified \textsc{rns} code to generate models of rotating neutron stars; besides the central energy density $\epsilon_c$, this requires supplying an axis ratio $\mathfrak{r} = r_p/r_e$ (where $r_p$ and $r_e$ are the polar and equatorial coordinate radii, respectively) which encodes the star's rotation rate. Internally, the \textsc{rns} code starts from a non-rotating configuration and then iteratively reduces the axis ratio up until the desired value is reached. As we are interested in the entire sequences from no rotation up to the Kepler limit, we have modified the \textsc{rns} code such that it reduces the axis ratio $\mathfrak{r}$ in small steps according to our needs and writes out the bulk quantities also of the intermediate solutions; this drastically reduces the computational expense of creating the database for our study. Equilibrium configurations are constructed on a $401 \times 801$ grid, demanding an accuracy of $10^{-7}$; the axis ratio at the mass-shedding limit is determined to $10^{-6}$. In total, we constructed 5058826 models of rotating neutron stars.

The conceptual simplicity of sequences with constant central energy density is one reason why they have been employed in several studies (e.g. \citealt{2004MNRAS.352.1089S, 2006MNRAS.368.1609D}). Moreover, several universal relations based on such sequences have been identified, including those involving the $f$-mode frequencies and damping times \citep{2008PhRvD..78f4063G, 2020PhRvL.125k1106K, 2011PhRvD..83f4031G}, as well as some for mass and radius~\citep{2022ApJ...934..139K}. In our study, we find that also other bulk quantities of neutron stars follow near universal behaviour along these sequences (this is not the case for other sequences such as those along which, e.g., the gravitational or baryon mass is held constant); in the following, we describe the data set on which we build the present study in more detail.

\section{Data set}
\label{sec:dataset}

Given the nine equation-of-state parametrisations presented in Sec.~\ref{sec:eos_params}, we generate a large number of equations of state by drawing random parameters from the uniform priors provided in Tabs.~\ref{tab:param-ppa}-\ref{tab:param_sos_annala}. In a first filtering step, we build large database of equations of state and discard those equations of state that fail to satisfy minimal physical and astrophysical constraints. Specifically, we require that the equation of state remains causal (i.e., $0 \le c_s^2 < 1$) up to the maximum-mass model, which must satisfy $M_\textrm{TOV} = 1.8\,M_\odot$; further, it should reach densities of at least $2n_{\rm sat}$ and if it reaches an energy density of $\epsilon = 1 \times 10^{15}\gperccc$, the star with this central density should be heavier than $0.5\,M_\odot$; last, the radius $R_{1.6}$ of a $1.6\,M_\odot$ star should lie between 9 and $14\,$km. Using these criteria, we generate equations of state for each parametrisation until 1000 satisfy the constraints; Tab.~\ref{tab:discarded} lists how many equations of state were needed for each parametrisation. The comparatively high rejection rate for the spectral representations arises because the corresponding power expansions (cf.~Sec.~\ref{ssec:spectral}) can produce rapidly increasing values of the equation-of-state quantities, which quickly exceed physically or astrophysically acceptable limits. In App.~\ref{app:m-r}, we show mass-radius diagrams for the 1000 randomly generated equations of state for each parametrisation, illustrating their characteristic differences.

This initial filtering ensures that the equations of state in our database are at least somewhat ``reasonable''; an astrophysically motivated data set, e.g., based on observational constraints, can then be obtained by further filtering from the database which we explain below. While this additional filtering step may appear unnecessary, it allows us to extract specific data sets without the need to newly generate equations of state, which is computationally considerably more expensive. We note that we explicitly do not include equations of state that feature a phase transition, even though these might lead to interesting phenomena in binary mergers (e.g.\ \citealt{2020PhRvL.124q1103W, 2020EPJST.229.3595B}); we expect that universal relations would take a more complicated shape when phase transitions are considered.

\begin{table}
    \centering
    \caption{Number of randomly generated equations of state required to obtain 1000 models that satisfy all the basic constraints listed in the first paragraph of Sec.~\ref{sec:dataset}.}
    \begin{tabular}{lr}
    \toprule
    Model & Count \\
    \midrule
    A20 &   \num{15334} \\
    G19 &    \num{6035} \\
    F19     &   \num{22317} \\
    PPA     &   \num{12644} \\
    GPPA    &  \num{106066} \\
    PB-S    &  \num{233308} \\
    HB-S    &  \num{106908} \\
    PB-C-S  & \num{2349389} \\
    HB-C-S  &  \num{230790} \\
    \bottomrule
    \end{tabular}
    \label{tab:discarded}
\end{table}

While the above mentioned conditions seem very lax in astrophysical terms, they do filter out a large number of randomly generated equations of state. For each equation of state, we pick two energy densities as integer multiples of \SI{0.05e15}{\gram\per\cubic\centi\metre}: one is the lowest density such that the corresponding neutron star has a mass of at least $0.7\,M_\odot$, the other one is the largest that is still on the stable branch (i.e., right before the maximum mass model). We then construct 10 sequences of constant central energy density between these two densities such that the masses of the non-rotating stars of one particular equation of state are roughly evenly spaced in their masses. The models along a sequence are evenly spaced in the axis ratio $\mathfrak{r}$ with $\Delta\mathfrak{r} = 0.01$; every sequence also contains the Keplerian model, which typically does not fit into the evenly spaced grid. With the axis ratio $\mathfrak{r}$ of the Keplerian model usually falling into the range of $\mathfrak{r}_K \in [0.50, 0.60]$, every sequence contains about 40-50 neutron star models, ranging from zero ($\Omega_n = 0$) to maximal ($\Omega_n = 1$) rotation. Starting from this fairly large database of sequences of stars, we may apply customised filters in order to arrive at a dataset ready for investigation.

To obtain an astrophysically reasonable data set, we discard all sequences of stars for which the corresponding equation of state violates one or more of the following observational constraints: the equation of state must support neutron stars with gravitational masses of at least $M \ge 1.97\,M_\odot$ \citep{2013Sci...340..448A}; the radius $R_{1.6}$ of a $1.6\,M_\odot$ neutron star must exceed $10.6\,\textrm{km}$ \citep{2017ApJ...850L..34B}, and the radius $R_{1.4}$ of a $1.4\,M_\odot$ star has to lie within the range $11.5\,\textrm{km} \le R_{1.4} \le 13.5\,\textrm{km}$ \citep{2021ApJ...918L..29R}; finally, the tidal deformability $\Lambda_{1.4}$ of a $1.4\,M_\odot$ neutron star must fall within the interval $120 \le \Lambda_{1.4} \le 800$ \citep{2018PhRvL.120q2703A, 2018PhRvL.121p1101A}. Furthermore, we discard all sequences whose non-rotating member is lighter than $1.05\,M_\odot$. This ensures that our universal relations are calibrated even for neutron stars that are slightly lighter than astrophysically expected \citep{2018MNRAS.481.3305S, 2015ApJ...812..143M}; with the only known exception being the strangely light $0.77^{+0.20}_{-0.17}\,M_\odot$ neutron star reported by \citet{2022NatAs...6.1444D}. Including such low-mass stars would also significantly weaken the universal relations derived in this study; more precisely, we observe a marked deterioration in their tightness when stars with masses $\lesssim 1\,M_\odot$ are included. We do not currently impose additional speed of sound constraints \citep{2018ApJ...860..149T, 2022ApJ...939L..34A}; merely the two parametrisations proposed by \citet{2019MNRAS.485.5363G} and \citet{2020NatPh..16..907A} explicitly incorporate them. The pQCD constraint $c_s^2 \rightarrow 1/3$ applies only at very high densities ($\approx 40n_0$), so deviations at lower densities do not constitute a significant violation of theoretical expectations.

\begin{figure}
    \centering
    \includegraphics[width=1\linewidth]{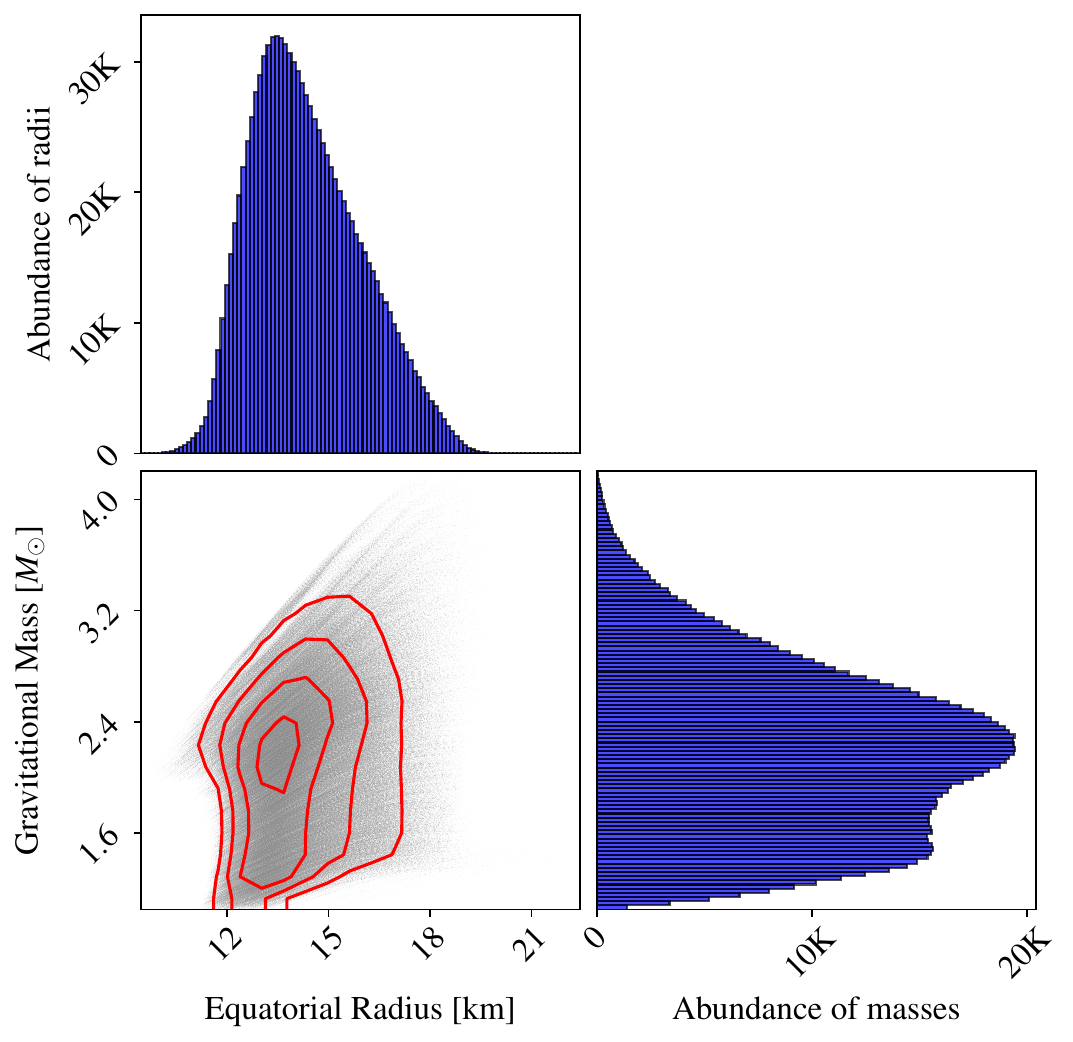}
    \caption{Corner plot of the dataset on which we construct the universal relations. Each black dot in the $M$ vs. $R_e$ diagram represents one of 998639 neutron star models in our dataset. Their density distribution is indicated by the red contour lines. The top and right diagram show the corresponding histograms of radius and mass using 100 bins, respectively.}
    \label{fig:mr_hist2d}
\end{figure}

After applying these filters to our database, each parametrisation yields a different number of ``valid'' equations of state. To avoid introducing an artificial bias, we restrict each parametrisation to 320 equations of state, corresponding to the minimum number obtained among all parametrisations. Our dataset then contains 21383 sequences of stars that comprise 998639 individual neutron star models. This is the dataset to which our universal relations will be calibrated. We visualise our dataset in Fig.~\ref{fig:mr_hist2d}.

Tighter and/or different constraints have been published, however, we intentionally keep our dataset broad as some of the published constraints show some tension and we intend to propose robust universal relations. Furthermore, as a test, we applied the constraints from \citet{2025PhRvX..15b1014K}, which are $R_{1.4} = 12.20^{+0.50}_{-0.48}\,\textrm{km}$ and $M_{\rm TOV} = 2.30^{+0.07}_{-0.20}\,M_\odot$ (but with no constraint on the tidal deformability $\Lambda$). These constraints are considerably tighter than those that we impose; in particular, the upper limit on $M_{\rm TOV}$ eliminates a large number of equations of state\footnote{Curiously, the two parametrisations PB-C-S and HB-C-S suffer the most from the upper limit on $M_{\rm TOV}$ and only roughly 100 (of initially 1000 in the entire database) equations of state survive these constraints; for other parametrisations, about 200-300 equations of state remain after filtering.}, resulting in a dataset (comprising 5355 sequences with 249486 neutron stars) that is roughly a quarter in size of the one we actually employ. Nonetheless, not only the universal relations remain largely unaltered (as they should) but also their relative errors are only marginally reduced when calibrating to this considerably more narrow dataset. While this demonstrates that the universal relations which we will discuss in Sec.~\ref{sec:urs} are robust, this also tells us that more recent astrophysical constraints still support a large variation in the properties of the equation of state.

\section{Universal relations}
\label{sec:urs}

The aim of this study is to provide universal relations for various bulk quantities of neutron stars, which allow to estimate to percent-level accuracy the value of a bulk quantity of a neutron star rotating with the angular rotation rate $\Omega$ using just the three numbers $M_\star$, $R_\star$, and $I_\star$ that belong to the non-rotating star of the corresponding sequence. In particular, we will propose universal relations for the following bulk quantities of a neutron star: gravitational mass $M$, baryon mass $M_0$, proper mass $M_p$, equatorial radius $R_e$, axis ratio $\mathfrak{r}$, moment of inertia $I$, effective compactness $\eta$\footnote{The effective compactness is defined as $\eta = \sqrt{\bar{M}^3 / I_{45}}$, with $\bar{M} := M / M_\odot$ and $I_{45} := I / 10^{45}\,\unit{g}\,\unit{cm}^2$.}, gravitational binding energy $W$, and the ratio of gravitational binding energy to rotational kinetic energy $T/W$. For each bulk quantity, two universal relations will need to be evaluated, which we will explain in the following.

\subsection{Normalised Bulk Quantities along Sequences}
\label{ssec:ur_sequences}

Let us now consider a bulk quantity $Q$. Along a given sequence, the value of this quantity evolves from that of the non-rotating star, $Q_\star = Q(\Omega = 0)$, to its value in the Keplerian model, $Q_K = Q(\Omega_K)$. We now normalise both the rotation rate and the quantity $Q$ such that both span the unit interval $[0, 1]$. The normalised quantities carry the subscript ``n'' and the linear mapping for $Q$ is
\begin{equation}
    Q_n := \frac{Q - Q_\star}{Q_K - Q_\star} \in [0, 1].
    \label{eq:def_Qn}
\end{equation}
We note that each bulk quantity $Q$ that we consider shows a monotonous behaviour along the sequences, which implies that the normalized quantity $Q_n$ indeed lies within $[0, 1]$. The angular rotation rate $\Omega$ is normalised in the same way, but since its non-rotating value vanishes, $\Omega_\star = 0$, the expression reduces to $\Omega_n := \Omega / \Omega_K$.

\begin{figure}
    \centering
    \includegraphics[width=1\linewidth]{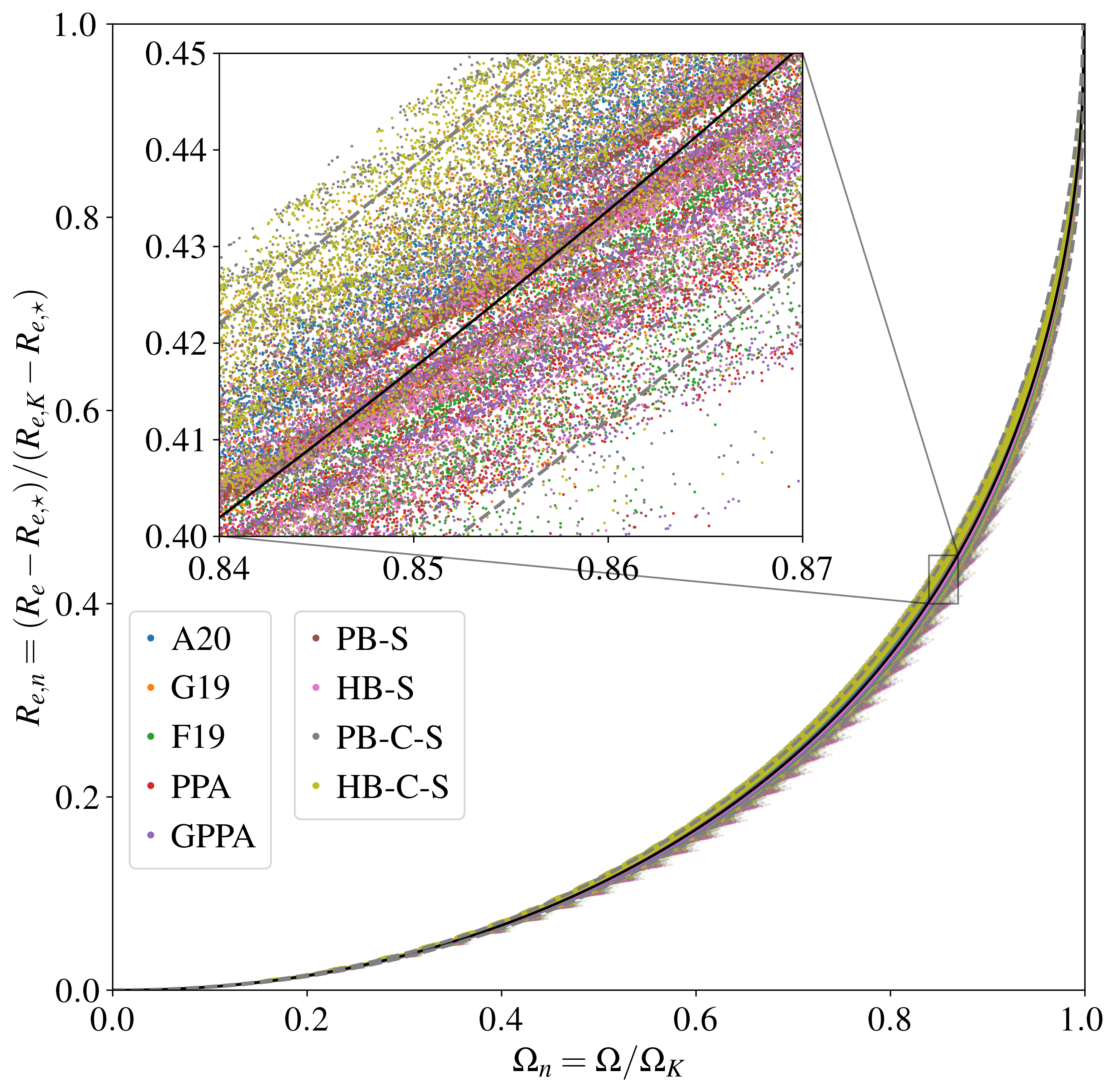}
    \caption{Scatter plot of the rescaled equatorial radius $R_{e,n}$ against the fractional angular rotation rate $\Omega_n$ for the 21383 sequences of stars in our dataset. As the graph shows a very large number of data points, we display an inset, in which we enlarge a small region of the graph; it becomes visible that the majority of the data points gather within $5\,\%$ (indicated by dashed grey lines, not shown in the main graph) of the polynomial fit (black solid line) and there are some with a larger deviation from the fit. The relative error for the equatorial radius $R_e$ is bounded by $1.3\,\%$; see the discussion in the main text for details.}
    \label{fig:scatter_R_e}
\end{figure}

As an example, we choose the bulk quantity $R_e$ and show the result of this normalisation in Fig.~\ref{fig:scatter_R_e}, which shows a scatter plot of the normalised equatorial radius $R_{e,n}$ against the normalised rotation rate $\Omega_n$. Different colours indicate different equation-of-state parametrisations. The normalisations of $R_e$ and $\Omega$ are such that the curves of any sequence will by construction pass through the points $(0, 0)$ and $(1, 1)$, which correspond to the non-rotating star and the star at the Kepler limit, respectively. In addition, we find that not only these two points belong to the graph of a sequence but, in fact, the entire graph connecting those two points is (with only small deviations) independent of the equation of state as well as the central energy density that is chosen for a particular sequence. We also find similar results for the other bulk quantities.

The figure strongly suggests a relation $Q_n = f(\Omega_n)$ and we propose a simple polynomial fit of the form
\begin{equation}
    f(\Omega_n) = \sum_{k=1}^4 c_{2k} \Omega_n^{2k},
    \label{eq:fit_quadratic}
\end{equation}
where the $c_{2k}$ are the parameters to be fitted. We omit the constant term to enforce $f(0) = 0$ and retain only even-powered terms, since the rotational corrections to all bulk quantities (at least those under consideration) are of quadratic order. This polynomial fit works very well for most of the bulk quantities except for the equatorial radius $R_e$ and the axis ratio $\mathfrak{r}$. Close to the Keplerian end of a sequence, the equatorial radius $R_e$ increases sharply as the material on the equator gets closer and closer to mass-shedding; this results in a very large gradient $\partial R_{e,n} / \partial \Omega_n$, which cannot be well resolved by our polynomial ansatz. A similar argument holds for the axis ratio $\mathfrak{r}$ as it is inversely proportional to the (coordinate) equatorial radius. For these two quantities, we use the fitting function
\begin{equation}
    f(\Omega_n) = 1 - \sqrt[\nu]{ 1 - \Omega_n^{\nu} } + \sum_{k=1}^4 c_{2k} \Omega_n^{2k},
    \label{eq:fit_quadratic_circ}
\end{equation}
which models the graph as a quadrant of a superellipse (featuring an infinite slope at $\Omega_n = 1$, i.e., at the Kepler limit) and catches the deviations from this superellipse with a polynomial ansatz as before. Here, the parameters to be fitted are the $c_{2k}$ and $\nu$.

Let us return to the example of the equatorial radius $R_e$ shown in Fig.~\ref{fig:scatter_R_e}. The black line represents the best fit based on the ansatz in Eq.~\eqref{eq:fit_quadratic_circ} (whose coefficients can be found in Tab.~\ref{tab:fitting2}), while the two dashed grey lines in the inset indicate a $5\,\%$ deviation from this fit. For this exemplary case of the equatorial radius, the vast majority of data points lie within this $5\,\%$ envelope around the best fit. This, however, is only the deviation of the normalised equatorial radius from the fitting function.

\begin{figure}
    \centering
    \includegraphics[width=1\linewidth]{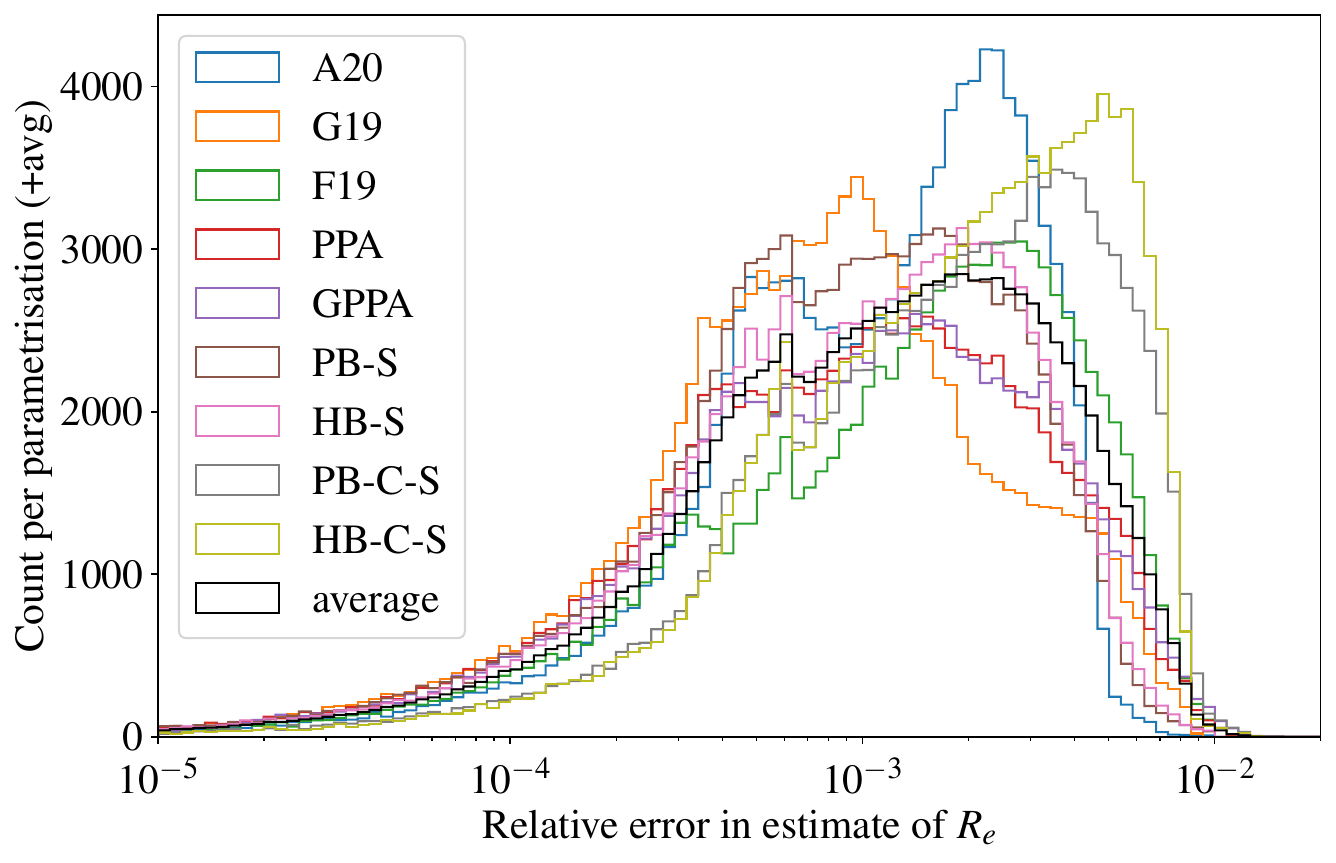}
    \caption{Histograms for the relative error in the estimate of the equatorial radius $R_e$ for the different parametrisations (in colour). The black line indicates the averaged histogram (the histogram of the entire dataset would scale by a factor of nine).}
    \label{fig:relerr_R_e}
\end{figure}

In general, for any bulk quantity $Q$, the unscaled value can be recovered from its normalised form using Eq.~\eqref{eq:def_Qn}, which yields
\begin{align}
    Q
    & = Q_\star + Q_n (Q_K - Q_\star)
    \\
    & = Q_\star + f(\Omega_n) (Q_K - Q_\star),
    \label{eq:Q_solved}
\end{align}
where the function $f(\Omega_n)$ is fitted individually for each bulk quantity. $Q_K$ is the Keplerian value of $Q$ which is hitherto unknown; we will discuss an estimate for it in Sec.~\ref{ssec:keplerian}. To quantify the accuracy of the fit, we evaluate the magnitude of the relative error
\begin{equation}
    RE_k = \left| \frac{Q_k - \hat{Q}_k}{Q_k} \right|,
    \label{eq:re_k}
\end{equation}
where $k$ is an index denoting the data points and $Q_i$ denotes the true value obtained from the \textsc{rns} code, and $\hat{Q}_i$ is the estimated value computed from the universal relation. For bulk quantities where the non-rotating value $Q_\star$ is nonzero, the relative error of the unscaled value $Q$ is typically smaller than that of its normalised corresponding normalised counterpart $Q_n$. In the exemplary case of the equatorial radius $R_e$, we find that the relative error of the estimated radii remains below $1.3\,\%$ across the entire dataset (this can be seen in Fig.~\ref{fig:relerr_R_e} where we show histograms for the relative error of $R_e$ for both the individual parametrisations as well as the entire data set). Moreover, as seen in Fig.~\ref{fig:scatter_R_e}, only a few data points deviate significantly from the fit, improving the radius estimate even further: the 99th percentile of the relative error is $0.7\,\%$, nearly half of the maximum relative error. The arithmetic mean of the relative error is just $0.18\,\%$, further indicating the excellent quality of the fit.

\begin{table*}
    \centering
    \caption{Fitting coefficients $c_k$ for the ansatz in Eq.~\eqref{eq:fit_quadratic} for several bulk quantities based on 998639 data points. Given the fitting coefficients, the data points are reconstructed using Eq.~\eqref{eq:Q_solved} and the absolute relative errors computed, $\textrm{RE} = |(Q_i - \hat{Q}_i)/Q_i|$. The last four columns show the mean relative error (Mean RE), the 95th and 99th percentile (P95 and P99), and the maximum relative error (Max RE).}
    \sisetup{
        table-format=1.6,
        table-number-alignment=center,
        separate-uncertainty = true,
        detect-all
    }
    \begin{tabular}{c|
        S[table-format=+1.4,round-mode=figures,round-precision=4]
        S[table-format=+1.4,round-mode=figures,round-precision=4]
        S[table-format=+1.5,round-mode=figures,round-precision=4]
        S[table-format=+1.5,round-mode=figures,round-precision=4]
        |
        S[table-format=1.2,table-space-text-post=\%,round-mode=places,round-precision=2]
        S[table-format=1.1,table-space-text-post=\%,round-mode=places,round-precision=1]
        S[table-format=1.1,table-space-text-post=\%,round-mode=places,round-precision=1]
        S[table-format=2.1,table-space-text-post=\%,round-mode=places,round-precision=1]}
    \toprule
    {Quantity} & {$c_2$} & {$c_4$} & {$c_6$} & {$c_8$} & {Mean RE} & {P95} & {P99} & {Max RE} \\
    \midrule
$M$ & 0.510789 & 0.485213 & -0.502541 & 0.506540 & 0.20\,\% & 0.58\,\% & 0.80\,\% & 2.29\,\% \\
$M_0$ & 0.505145 & 0.489682 & -0.509044 & 0.514217 & 0.21\,\% & 0.59\,\% & 0.81\,\% & 2.29\,\% \\
$M_p$ & 0.508623 & 0.483055 & -0.495497 & 0.503819 & 0.20\,\% & 0.59\,\% & 0.81\,\% & 2.32\,\% \\
$T/W$ & 0.752448 & 0.252474 & -0.223819 & 0.218898 & 1.13\,\% & 3.36\,\% & 4.59\,\% & 10.60\,\% \\
$I$ & 0.403396 & 0.660051 & -0.872718 & 0.809271 & 0.66\,\% & 1.93\,\% & 2.77\,\% & 8.49\,\% \\
$\eta$ & 0.403909 & 0.435953 & -0.072576 & 0.232713 & 0.06\,\% & 0.19\,\% & 0.27\,\% & 0.55\,\% \\
$W$ & 0.491907 & 0.489277 & -0.504384 & 0.523200 & 0.36\,\% & 1.09\,\% & 1.58\,\% & 4.43\,\% \\
    \bottomrule
    \end{tabular}
    \label{tab:fitting1}
\end{table*}

\begin{table*}
    \centering
    \caption{Fitting coefficients $\nu$ and $c_k$ for the ansatz in Eq.~\eqref{eq:fit_quadratic_circ} for the axis ratio $\mathfrak{r}$ and the equatorial radius $R_e$ based on 998639 data points. The last four columns are the same as in Tab.~\ref{tab:fitting1}.}
    \sisetup{
        table-format=1.6,
        table-number-alignment=center,
        separate-uncertainty = true,
        detect-all
    }
    \begin{tabular}{c|
        S[table-format=+1.3,round-mode=figures,round-precision=4]
        S[table-format=+1.5,round-mode=figures,round-precision=4]
        S[table-format=+1.4,round-mode=figures,round-precision=4]
        S[table-format=+1.4,round-mode=figures,round-precision=4]
        S[table-format=+1.6,round-mode=figures,round-precision=4]
        |
        S[table-format=1.2,table-space-text-post=\%,round-mode=places,round-precision=2]
        S[table-format=1.1,table-space-text-post=\%,round-mode=places,round-precision=1]
        S[table-format=1.1,table-space-text-post=\%,round-mode=places,round-precision=1]
        S[table-format=2.1,table-space-text-post=\%,round-mode=places,round-precision=1]}
    \toprule
    {Quantity} & {$\nu$} & {$c_2$} & {$c_4$} & {$c_6$} & {$c_8$} & {Mean RE} & {P95} & {P99} & {Max RE} \\
    \midrule
$\mathfrak{r}$ & 1.681203 & -0.129786 & 0.554007 & -0.553104 & 0.127180 & 0.35\,\% & 1.02\,\% & 1.39\,\% & 2.35\,\% \\
$R_e$ & 2.178385 & 0.031602 & -0.127447 & 0.078108 & 0.019801 & 0.18\,\% & 0.54\,\% & 0.74\,\% & 1.31\,\% \\
    \bottomrule
    \end{tabular}
    \label{tab:fitting2}
\end{table*}

All fitting coefficients and relative errors (plus the 95th percentile) for the equatorial radius and the axis ratio (which are fitted using the ansatz in Eq.~\eqref{eq:fit_quadratic_circ}) can be found in Tab.~\ref{tab:fitting2}, while the coefficients and error quantifiers for those bulk quantities which are modelled using the ansatz in Eq.~\eqref{eq:fit_quadratic} are listed in Tab.~\ref{tab:fitting1}. The sum of the coefficients $c_i$ obtained from the least-squares fits to Eq.~\eqref{eq:fit_quadratic} deviates from 1.0 by up to 0.0031. We therefore rescale them linearly so that their sum is exactly 1.0, ensuring that the fit passes through the point $(1.0, 1.0)$ as intended.

In order to visualise the magnitude of the relative errors of the final quantity in our example, i.e. of the equatorial radius $R_e$, we show them in Fig.~\ref{fig:relerr_R_e} as histograms. The different colours indicate the individual equation-of-state parametrisations, while the black line represents the ``averaged histogram''. We opt for the averaged histogram since the shape is more relevant than the absolute counts; the histogram of the combined dataset would merely scale the height by a factor of nine. While the histograms all peak at a relative error of a few per mille, they are a bit different in detail. This is owing to the fact that each parametrisation introduces some kind of bias as we will discuss in the following section.

\subsection{Discussion of the URs for the Normalised Quantities}

All discussed parametrisations can straightforwardly be extended to an arbitrary number of parameters, and hence faithfully represent any equation of state (at least those with no phase transition); this is also true for the speed-of-sound-parametrisation G19 (cf.~Sec.~\ref{sec:sos-greif}) to which an arbitrary number of parametrised Gaussians can be added (instead of only one). However, with the number of parameters being artificially limited to a small number (and yet realistic equation of state can be approximated to very good precision), each parametrisation is naturally biased in a slightly different way.

\begin{figure}
    \centering
    \includegraphics[width=1\linewidth]{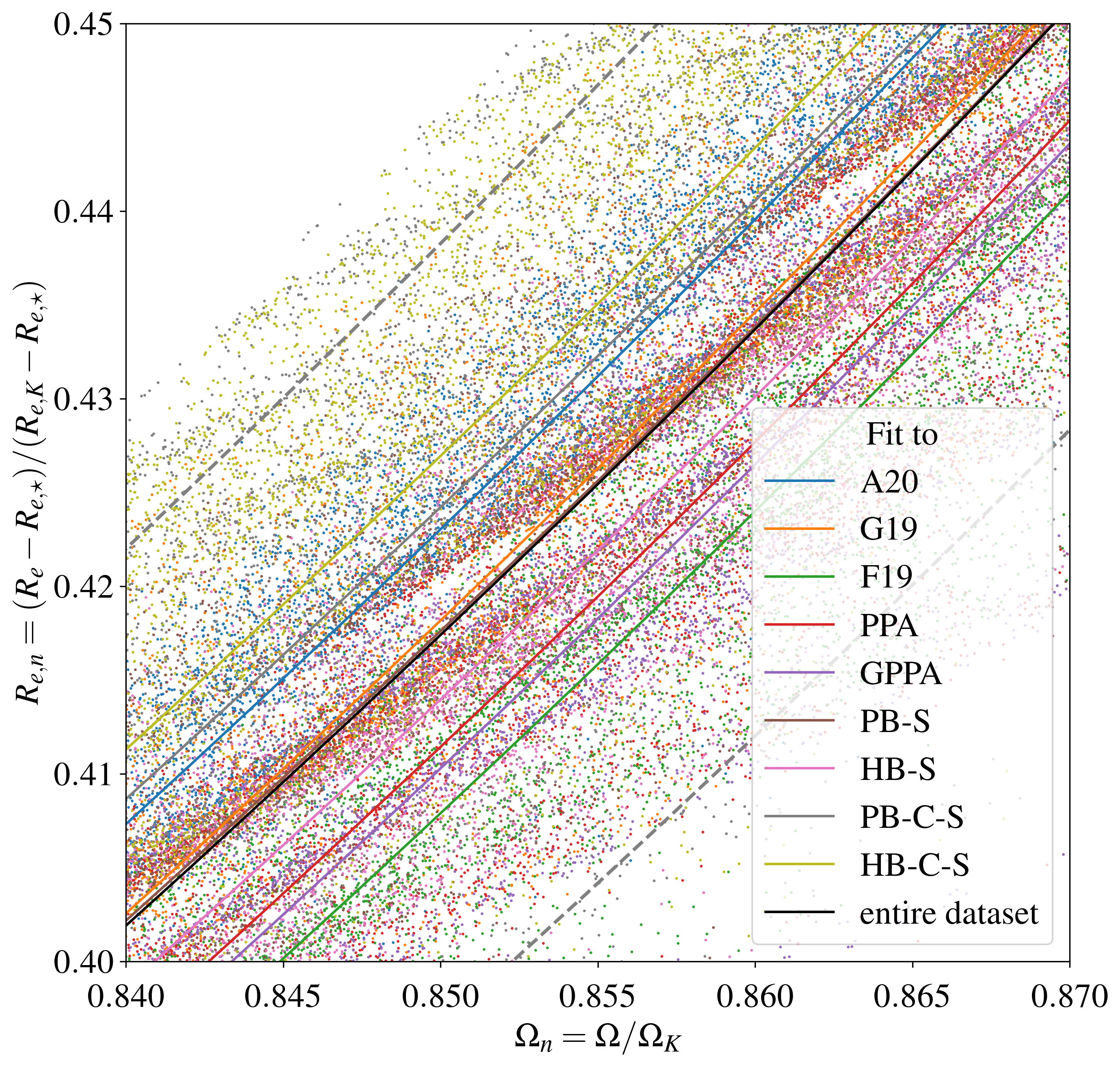}
    \caption{The inset of Fig.~\ref{fig:scatter_R_e} for the equatorial radius $R_e$ including the polynomial fits (colored solid lines) for each parametrisation. It is obvious that each parametrisation introduces its own bias; the black solid line indicates the fit to the entire dataset and the grey dashed lines indicate a $5\,\%$ deviation from this fit.}
    \label{fig:scatter_R_e_inset}
\end{figure}

Figure~\ref{fig:scatter_R_e_inset} shows the inset of Fig.~\ref{fig:scatter_R_e} in which the coloured solid lines indicate the best fits for the individual parametrisations. While these are very close to each other, they do deviate from one another due to the parametrisation bias (we do not report the corresponding coefficients as this would result in massive tables with little purpose). We do not observe a clear pattern among the individual fits when comparing them for different bulk quantities such as, e.g., an ordering or relative deviations; hence, we do not attempt to explain the impact of a particular parametrisation on the different bulk quantities. Instead, we point out that, even when fitting the data of all parametrisations collectively, the universal relations possess very high accuracy: The mean of the relative errors in our dataset is less than $1.13\,\%$ as can be seen in Tabs.~\ref{tab:fitting1} and \ref{tab:fitting2}. These tables also show that the 99th percentile is considerably better than the maximum relative error.

We have singled out those neutron stars models in our dataset for which the universal relation for some bulk quantity results in a large relative error; usually, many other members of the sequence to which they belong, suffer from a similarly large relative error for this particular bulk quantity. However, it is important to note that other bulk quantities of such neutron star models (or their sequence) are not ``outliers'' as well but usually are estimated to much higher accuracy. In other words, it is not the case that certain sequences are generally badly estimated, but it is just one particular bulk quantity along one sequence that may suffer from estimates of lower accuracy. We are at this time not able to detect a pattern as to what these ``outliers'' might have in common; however, we note that the 99th percentile of the relative errors is at percent-level accuracy for all bulk quantities (with the exception of $T/W$ for which the 99th percentile is slightly larger at $4.6\,\%$), and the mean relative error is considerably better, resulting generally in very good estimates for any of the bulk quantities.

We briefly comment on the universal relations for $I$ and $T/W$ that we reported in our prior study \citep{2023PhRvD.108l4056K}. There, we reported maximum relative errors of $1.8\,\%$ and $4.3\,\%$ for these two bulk quantities, respectively, which is considerably better than the maximum relative errors of $8.5\,\%$ and $10.6\,\%$ that we report in the present study. This is simply owing to the fact that the prior dataset was considerably smaller, as it was based merely on piecewise-polytropic approximations of realistic equations of state. In the present study, we lift both limitations: we consider any equation of state that satisfies a set of astrophysical constraints and average over nine different equation-of-state parametrisations. With this in mind, it is noteworthy that the universal relations still allow to estimate these quantities to very good accuracy (in particular when considering the mean relative error).

Furthermore, we report that the angular momentum $J$ and the rotational kinetic energy $T$ have a dramatically wider spread in their scatter plots, where the maximum relative errors would be $39\,\%$ and $37\,\%$, respectively. Hence, we do not report coefficients for universal relations for these quantities but instead suggest to estimate them via $J = I\Omega$ and $T=\frac{1}{2} I\Omega^2$. Similarly, the normalised angular momentum $a := J/M^2 = I\Omega / M^2$ is best estimated as a combined quantity, too.

\subsection{Keplerian Values of the Bulk Quantities}
\label{ssec:keplerian}

In the previous section, we discussed how the curves of one given bulk quantity along sequences of constant central energy density are nearly the same when normalising them to the unit square. The universality can be used to estimate the bulk quantity $Q$ of a star with the rotation rate $\Omega$ via the expression in Eq.~\eqref{eq:Q_solved}, i.e., $Q = Q_\star + f(\Omega_n) (Q_K - Q_\star)$. However, we obviously need to know $Q_\star$, $Q_K$, and $\Omega_n$ to evaluate this formula. The value of $Q_\star$ is easily accessible by solving the TOV and Hartle's equation \citep{1967ApJ...150.1005H} (and perhaps, e.g., for the baryon mass, calculating an integral over some of the solution functions). For $\Omega_n$, we need to know $\Omega_K$ since $\Omega_n = \Omega / \Omega_K$. A straightforward idea would be to employ the \textsc{rns} code to iterate for the Kepler model of that sequence and determine $\Omega_K$ and $Q_K$ this way. But this wouldn't make sense procedurally as one could simply iterate for the desired rotation rate instead of the Kepler model and obtain the correct value for $Q$ directly.

\begin{table*}
\centering
\caption{Fitting coefficients $d_k$ to estimate Keplerian values of bulk quantities. The fitting function is provided in Eq.~\eqref{eq:fit_kepler} and the fit is based on 21383 data points. For each quantity $Q$ (first column), the corresponding combinations for $x$ and $y$ that are used in Eq.~\eqref{eq:fit_kepler} are listed in columns 2 and 3. The last four columns are the same as in Tab.~\ref{tab:fitting1}. We include an additional fit for the angular rotation rate $\Omega$ based on $x_\Omega = C_\star$ for comparison with previously published results by \protect\cite{2022ApJ...934..139K}; see also~Sec.~\ref{sec:comparison}}.
\sisetup{
  table-format=1.6,
  table-number-alignment=center,
  separate-uncertainty = true,
  detect-all
}
\resizebox{\linewidth}{!}{%
\begin{tabular}{l|ll|
                S[table-format=+1.5,round-mode=figures,round-precision=4]
                S[table-format=+1.5,round-mode=figures,round-precision=4]
                S[table-format=+1.5,round-mode=figures,round-precision=4]
                S[table-format=+1.6,round-mode=figures,round-precision=4]
                S[table-format=+1.8,round-mode=figures,round-precision=4]
                |
                S[table-format=1.2,table-space-text-post=\%,round-mode=places,round-precision=2]
                S[table-format=1.1,table-space-text-post=\%,round-mode=places,round-precision=1]
                S[table-format=1.1,table-space-text-post=\%,round-mode=places,round-precision=1]
                S[table-format=2.1,table-space-text-post=\%,round-mode=places,round-precision=1]}
\toprule
Q & $x_Q$ & $y_Q$ 
& {$d_0$} & {$d_1$} & {$d_2$} & {$d_3$} & {$d_4$} 
& {Mean RE} & {P95} & {P99} & {Max RE} \\

\midrule
$\mathfrak{r}$ & $I_\star R_\star^{-3} C_\star^{2}$ & $\mathfrak{r}_K \cdot C_\star^{-5/2}$ & 1.20408325 & 0.29822885 & 0.19013019 & 0.01509368 & 0.00045007 & 0.34\,\% & 0.84\,\% & 1.40\,\% & 4.1\,\% \\
$M$ & $I_\star M_\star^{-3} C_\star^{-2}$ & $M_K \cdot M_\star^{-1} C_\star^{-7/2}$ & -2.87328566 & 2.95439426 & -0.44062142 & 0.04489152 & -0.00177531 & 0.33\,\% & 0.81\,\% & 1.45\,\% & 4.2\,\% \\
$M_0$ & $I_\star M_\star^{-3} C_\star^{-2}$ & $M_{0,K} \cdot M_\star^{-1} C_\star^{-7/2}$ & -2.49045968 & 2.91590241 & -0.44588390 & 0.04574362 & -0.00180352 & 0.77\,\% & 1.94\,\% & 2.68\,\% & 4.1\,\% \\
$M_p$ & $I_\star R_\star^{-3} C_\star^{2}$ & $M_{p,K} \cdot R_\star^{-1} C_\star^{2}$ & -0.65173884 & 0.10516364 & -0.16209410 & -0.01487131 & -0.00054341 & 0.35\,\% & 0.89\,\% & 1.68\,\% & 4.7\,\% \\
$R_e$ & $I_\star R_\star^{-3} C_\star^{3}$ & $R_{e,K} \cdot R_\star^{-1} C_\star$ & -0.75540873 & -0.16383734 & -0.05081980 & -0.00319464 & -0.00008408 & 0.36\,\% & 0.75\,\% & 1.06\,\% & 2.6\,\% \\
\midrule
$(T/W)$ & $I_\star M_\star^{-3}$ & $(T/W)_K \cdot C_\star^{-9/2}$ & -8.57157958 & 14.02466769 & -6.27489138 & 1.55115869 & -0.14617660 & 0.84\,\% & 1.59\,\% & 2.20\,\% & 4.5\,\% \\
$T$ & $I_\star R_\star^{-3} \sqrt{C_\star}$ & $T_K \cdot R_\star^{-1} C_\star^{4}$ & -2.35666622 & 2.32252703 & 0.09478001 & 0.09242333 & 0.00973075 & 1.54\,\% & 3.97\,\% & 5.53\,\% & 9.8\,\% \\
$W$ & $I_\star C_\star^{-4}$ & $W_K \cdot C_\star^{-3}$ & 50.04919450 & -18.38958725 & 2.64240136 & -0.16451241 & 0.00381160 & 1.21\,\% & 3.14\,\% & 4.23\,\% & 8.7\,\% \\
$I$ & $I_\star R_\star^{-3}$ & $I_K \cdot R_\star^{-3} C_\star^{3}$ & -1.31022110 & 2.04566361 & 0.58754667 & 0.35822250 & 0.04599738 & 1.65\,\% & 4.01\,\% & 5.72\,\% & 10.2\,\% \\
$\eta$ & $I_\star M_\star^{-3}$ & $\eta_K \cdot C_\star^{-1/2}$ & -0.01929515 & -0.05802558 & -0.02719570 & 0.00089426 & -0.00008745 & 0.34\,\% & 0.81\,\% & 1.10\,\% & 1.6\,\% \\
\midrule
$\Omega$ & $I_\star M_\star^{-1} R_\star^{-2}$ & $\Omega_K \cdot R_\star C_\star^{-1/2}$ & -7.03846524 & -0.10002603 & -1.08559022 & -1.07599290 & -0.33464023 & 0.14\,\% & 0.34\,\% & 0.45\,\% & 1.0\,\% \\
$\Omega$ & $I_\star C_\star$ & $\Omega_K \cdot R_\star C_\star^{-1/2}$ & -5.83370868 & 2.84806903 & 2.20553484 & 0.78226645 & 0.10432794 & 0.55\,\% & 1.31\,\% & 1.92\,\% & 3.4\,\% \\
$J$ & $I_\star R_\star^{-3} \sqrt{C_\star}$ & $J_K \cdot R_\star^{-2} C_\star^{4}$ & -2.03538030 & 1.41674857 & -0.17141823 & 0.03899650 & 0.00564537 & 1.78\,\% & 4.26\,\% & 5.96\,\% & 9.9\,\% \\
$a$ & $I_\star M_\star^{-3} C_\star^{-1}$ & $a_K \cdot C_\star^{-7/2}$ & -2.13009962 & 3.21238258 & -0.60716557 & 0.08943769 & -0.00500314 & 0.62\,\% & 1.14\,\% & 1.46\,\% & 2.5\,\% \\

\bottomrule
\end{tabular}
}
    \label{tab:fitting3}
\end{table*}

As in the previous work \citep{2023PhRvD.108l4056K}, we avoid using the \textsc{rns} code and instead propose additional universal relations that allow the Keplerian value $Q_K$ of the bulk quantity $Q$ to be estimated from the triple $(M_\star, R_\star, I_\star)$---that is, the mass, radius and moment of inertia of the corresponding non-rotating neutron star with the same central energy density. These three quantities can be computed efficiently by solving the TOV equation and Hartle's slow-rotation formalism \citep{1967ApJ...150.1005H}. Several authors have proposed universal relations to estimate the moment of inertia $I_\star$ from just $M_\star$ and $R_\star$ (see, e.g., \citealt{2002A&A...396..917B, 2005ApJ...629..979L, 2016MNRAS.459..646B}); however, these relations can lead to relative errors of up to $10\,\%$. Since this is larger than the accuracy we aim for---and as we were unable to identify relations that considerably improve the estimate for $I_\star$---we treat $I_\star$ as an ``independent'' input that carries information in addition to $M_\star$ and $R_\star$. We then construct suitable combinations of the triple $(M_\star, R_\star, I_\star)$ and the target quantity $Q_K$ that yield accurate estimates of the Keplerian value.

\begin{figure}
    \centering
    \includegraphics[width=1\linewidth]{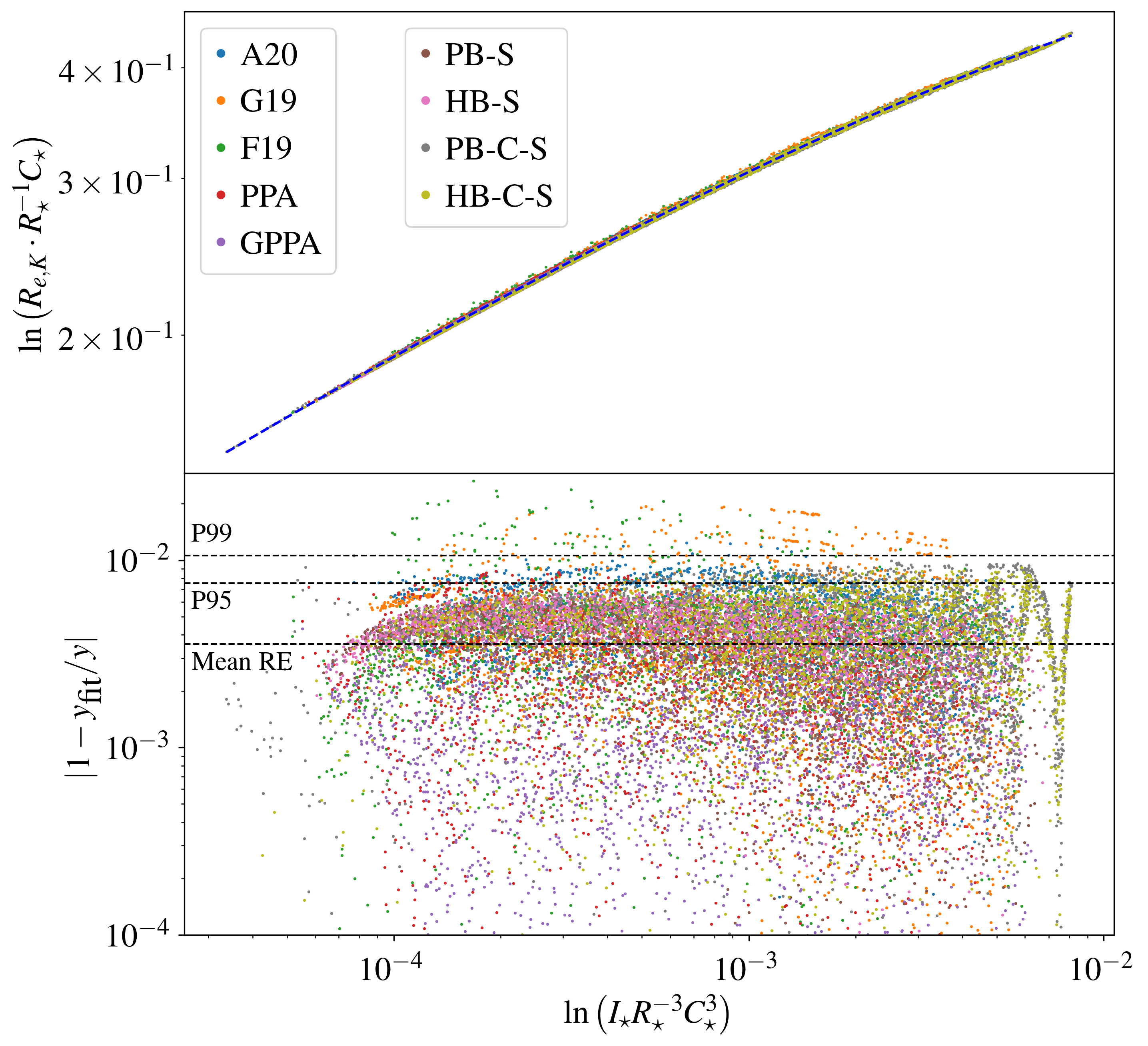}
    \caption{The scaled equatorial radius at the Kepler limit against the chosen combination of mass, radius, and moment of inertia of the non-rotating neutron star with the same central energy density as the rotating stars. The bottom panel shows the relative deviation of the fit (black solid line) from the true values. The maximum relative error is $2.6\,\%$, and the mean relative error (Mean RE) and its 95th (P95) and 99th percentile (P99) are indicated by dashed lines.}
    \label{fig:scatter_Kepler_R_e}
\end{figure}

We now apply this approach to our exemplary case from earlier, where we consider the equatorial radius $R_e$ as bulk quantity. A suitable combination of $(M_\star, R_\star, I_\star)$ and $R_{e,K}$ is
\begin{equation}
    x_{R_e} = I_\star R_\star^{-3} C_\star^3
    \quad\text{and}\quad
    y_{R_e} = R_{e,K} R_\star^{-1} C_\star,
\end{equation}
where we have introduced the compactness $C_\star := M_\star / R_\star$ as an abbreviation. In general, we use only powers of $M_\star$, $R_\star$, and $I_\star$ for the $x$-variable, and the $y$-variable contains the Keplerian value $Q_K$ multiplied with powers of $M_\star$ and $R_\star$. We show in the top half of Fig.~\ref{fig:scatter_Kepler_R_e} a scatter plot of the variables $\ln x_{R_e}$ and $\ln y_{R_e}$ for the equatorial radius $R_e$.

Again, the graph strongly motivates introducing a fitting function between $x_{R_e}$ and $y_{R_e}$. Here, for the Keplerian values of some bulk quantity $Q$, we take a general polynomial ansatz of the form
\begin{equation}
    \ln y_Q = \sum_{k=0}^4 d_k (\ln x_Q)^k.
    \label{eq:fit_kepler}
\end{equation}
The coefficients of the best fit for various bulk quantities can be found in Tab.~\ref{tab:fitting3}. Note that the combined quantities $x_Q$ and $y_Q$ contain very different combinations of $M_\star$, $R_\star$, $I_\star$, and the Keplerian quantity $Q_K$, and hence it is not advisable to compare the coefficients $d_k$ across the different bulk quantities. The particular combinations for $x_Q$ and $y_Q$ listed in the table have been found by systematically varying the exponents of $M_\star$, $R_\star$, and $I_\star$ until the corresponding universal relations turned out sufficiently tight.

\begin{figure}
    \centering
    \includegraphics[width=1\linewidth]{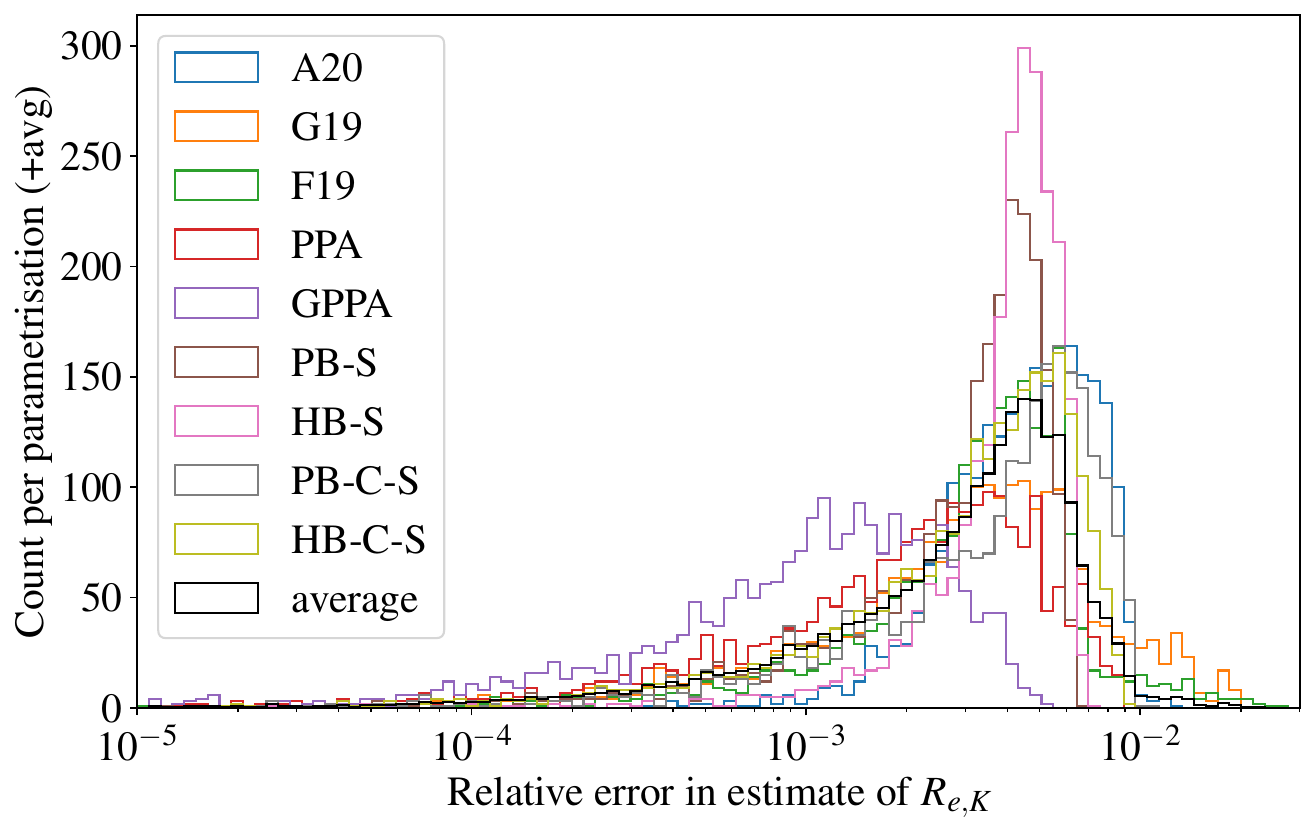}
    \caption{Histograms for the relative error in the estimate of the Keplerian value $R_{e,K}$ for the different parametrisations (in colour). The black line indicates the averaged histogram. The maximum relative error is $2.6\,\%$, while the 99th percentile is $1.1\,\%$.}
    \label{fig:relerr_R_e_K_hist}
\end{figure}

In Fig.~\ref{fig:scatter_Kepler_R_e}, we show the fitting function in the top graph with a black line. The lower half of the graph shows the relative error when estimating $R_{e,K}$ purely using $M_\star$, $R_\star$, and $I_\star$. We see that the maximum relative error across 21383 data points is $2.6\,\%$. Only a few outliers are estimated to worse than $1\,\%$ accuracy: the 99th percentile is $1.1\,\%$. The average mean of the relative errors is $0.36\,\%$. These numbers tell us that the equatorial radius $R_{e,K}$ of the Keplerian model can be very precisely estimated from the triple $(M_\star, R_\star, I_\star)$ of the corresponding non-rotating star. We also show histograms of the relative errors for each individual parametrisation in Fig.~\ref{fig:relerr_R_e_K_hist}.

The last ingredient that is missing to evaluate Eq.~\eqref{eq:Q_solved} is the angular rotation rate $\Omega_K$ of the Keplerian model. In the same way as for all bulk quantities, we propose suitable combinations for $x_\Omega$ and $y_\Omega$, which for $\Omega_K$ are
\begin{equation}
    x_\Omega = I_\star M_\star^{-1} R_\star^{-2}
    \quad\text{and}\quad
    y_\Omega = \Omega_K R_\star C_\star^{-1/2}.
\end{equation}
With the corresponding coefficients as listed in Tab.~\ref{tab:fitting3}, $\Omega_K$ can be estimated with an average discrepancy of only $0.14\,\%$, while the worst relative error in our dataset is $1.0\,\%$. For some sequences, the estimate of $\Omega_K$ is smaller than its true value, allowing $\Omega_n = \Omega / \Omega_K$ in Eq.~\eqref{eq:Q_solved} to exceed 1.0; in this case, we set $\Omega_n = 1.0$. \citet{2022ApJ...934..139K} proposed a similar relation for $\Omega_K$; however, in that case without employing the information contained in $I_\star$, i.e., the authors chose $x_\Omega = C_\star$. For the sake of comparison, we provide a fit for this choice of variable as well; we find that as a function of the compactness only, $\Omega_K$ can be estimated to better than $3.4\,\%$ accuracy, while the average discrepancy is $0.55\,\%$; however, this example demonstrates that the additional information carried in $I_\star$ may considerably improve the estimate. Even without admixing $I_\star$, the Kepler velocity can be estimated very well. Keep in mind that these universal relations are very tight, even though they are generated using random equations of state based on nine different equation-of-state parametrisations.

As we did earlier, we define a relative error
\begin{equation}
    RE_{Q,k} = \left| \frac{y_{Q,k} - \hat{y}_{Q,k}}{y_{Q,k}} \right|,
\end{equation}
where $y_{Q,k}$ denotes the true value of the compound variable $y_Q$ and $\hat{y}_{Q,k}$ is its estimate from the universal relation (as before, $k$ is an index denoting the data points). For the bulk quantities considered in this study, we show the fitting coefficients, maximum and mean relative error as well as the 95th and 99th percentile in Tab.~\ref{tab:fitting3}.

\subsection{Discussion of the URs for the Keplerian Values}

The maximum relative errors for the Keplerian estimates shown in Tab.~\ref{tab:fitting3} are slightly larger than those for the sequences of rotating stars. However, we find again, that the 99th percentile or the mean relative error are considerably smaller than the maximum relative error which indicates that the Keplerian value $Q_K$ of a bulk quantity may be very well estimated from the triple $(M_\star, R_\star, I_\star)$.

\begin{figure}
    \centering
    \includegraphics[width=1\linewidth]{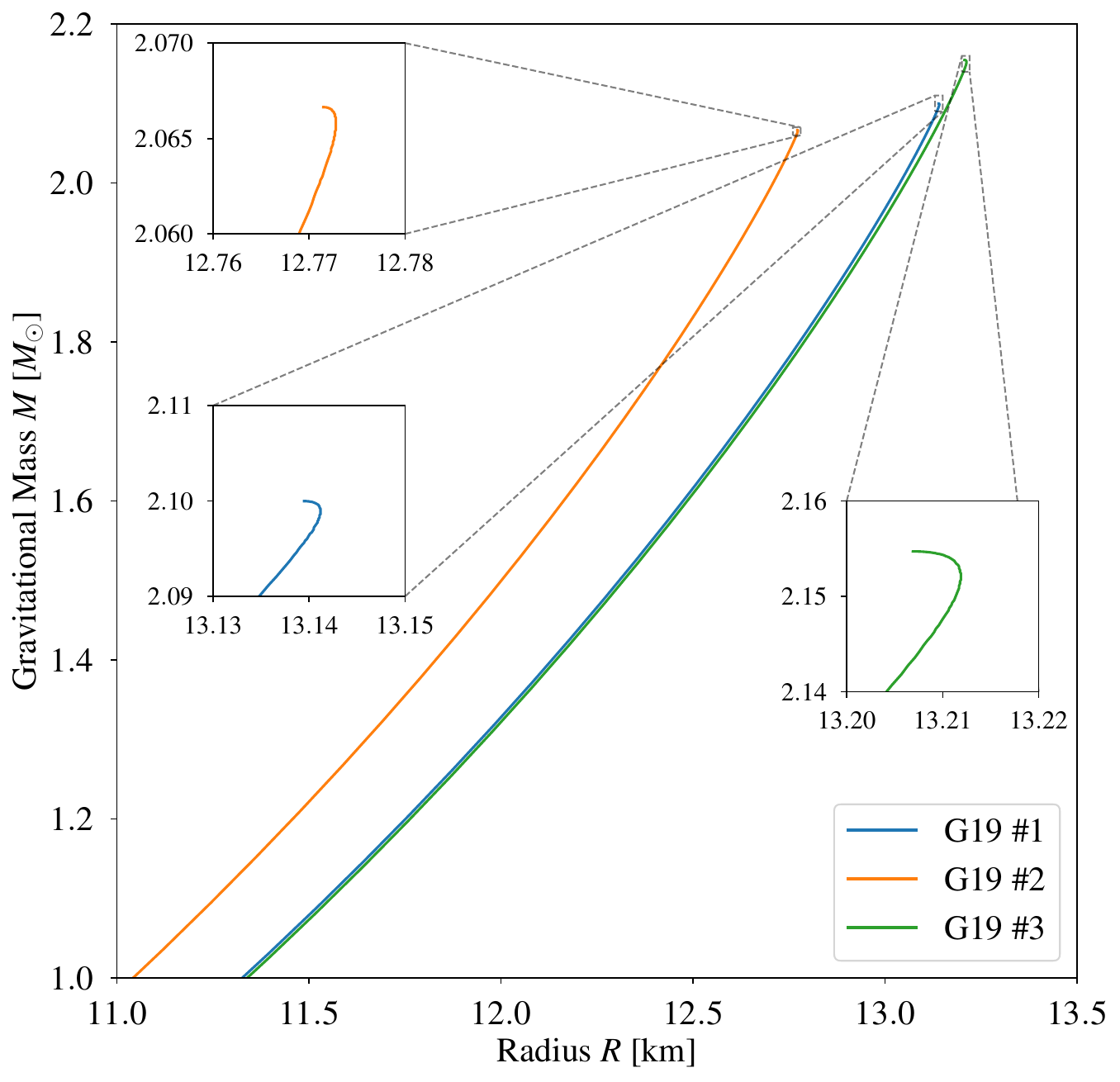}
    \caption{Mass-radius curves of three equations of state from the G19 parametrisation for which the estimate of the equatorial radius $R_e$ of the Keplerian model has the largest deviation from the true values (cf. the orange data points in Fig.~\ref{fig:scatter_Kepler_R_e}). The insets show the vicinity of the maximum mass models which are only marginally different from the models with maximum radius.}
    \label{fig:mr_sosgreif}
\end{figure}

Similar to the universal relations for the sequences of stars (cf.~Sec.~\ref{ssec:ur_sequences}), we investigated for which sequences the estimate of the Keplerian value of a bulk quantity suffers from larger relative errors; again, it proves difficult to find a conclusive pattern. In some cases, we find that sequences belonging to the same equation of state have similarly large relative errors in their Keplerian values, indicating that this particular equation of state has properties that lead to less accurate estimates; indeed, the worst relative errors from the G19 parametrisation (orange data points) in Fig.~\ref{fig:scatter_Kepler_R_e} stem from equations of state that satisfy our imposed astrophysical constraints but look somewhat unusual in the sense that mass and radius of the two models with maximum mass and with maximum radius are only marginally different from each other. To clarify this, we show the mass-radius curves of the three equations of state with the largest relative errors in Fig.~\ref{fig:mr_sosgreif}. A common feature of these three particular equations of state is that their Gaussians have a large amplitude, are rather narrow (with $a_3 \lesssim 0.3$) and are located at a fairly low density (with $a_2 \lesssim 2$). While such a speed-of-sound profile may also be produced by other parametrisations, the G19 parametrisation is particularly prone to this behaviour.

We show the coefficients to Eq.~\eqref{eq:fit_kepler} for various bulk quantities in Tab.~\ref{tab:fitting3}. We also include coefficients for the angular momentum $J$, the normalised angular momentum $a = J/M^2$, and the rotational kinetic energy $T$ for which we do not propose universal relations for the sequences: Their Keplerian values can be estimated to good accuracy while the sequences have a very large spread (as explained above). Finally, we also provide coefficients for the Keplerian rotation rate $\Omega_K$ as this quantity is necessary for calculating $\Omega_n$.

\section{Comparison to Previous Work}
\label{sec:comparison}

Universal relations concerned with rotating configurations in general or with Keplerian models in particular have been under investigation in the past. We are going to present comparisons of our present work to previous work where a meaningful comparison is possible.

\begin{figure}
    \centering
    \includegraphics[width=1\linewidth]{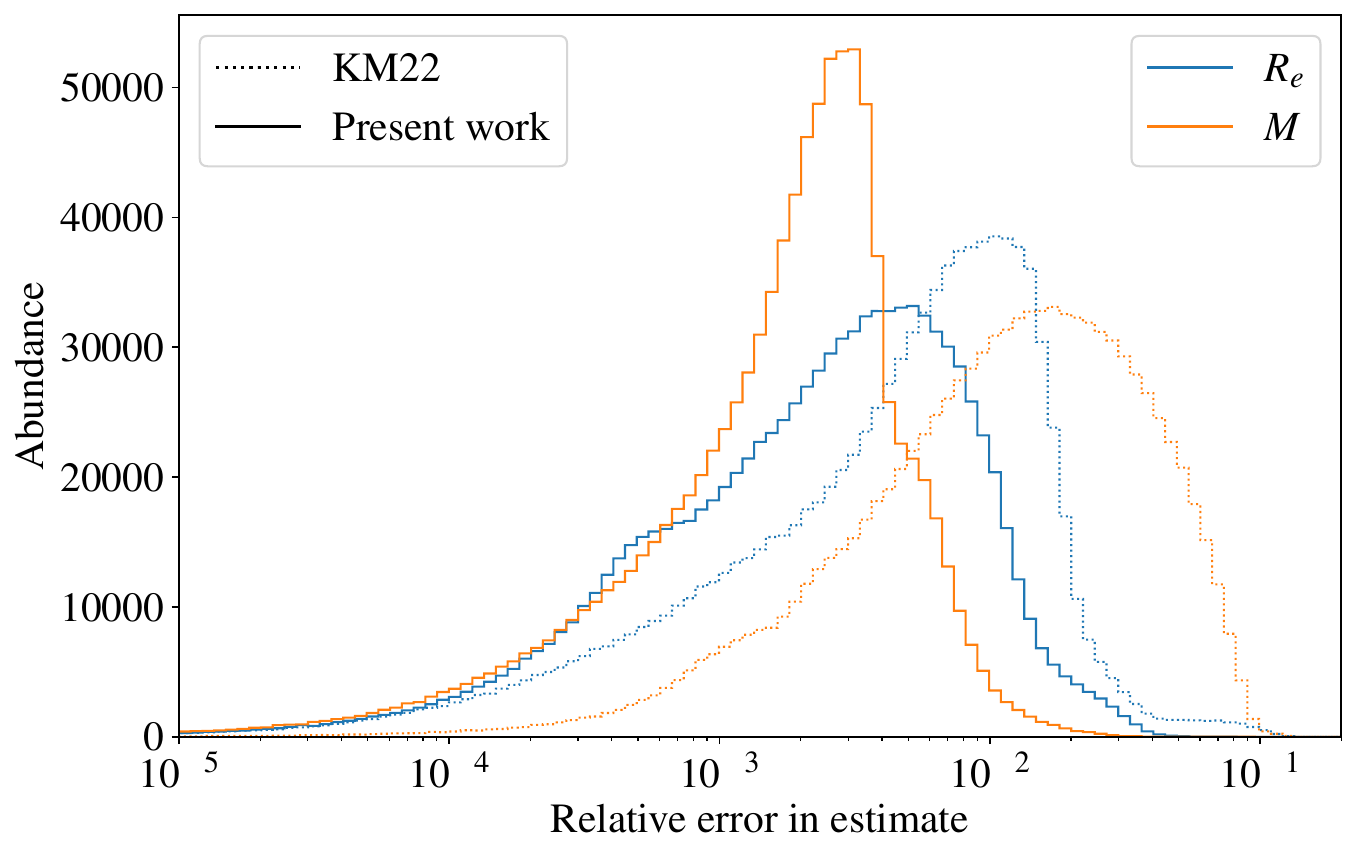}
    \caption{Histograms of the relative errors in the gravitational mass $M$ (orange) and equatorial radius $R_e$ (blue) obtained from the universal relations of~\protect\cite{2022ApJ...934..139K} (dotted lines) and the relations derived in this work (solid lines). The new relations yield a modest improvement for the radius estimate and an approximately one-order-of-magnitude improvement for the mass estimate on average. For both quantities, the maximum relative errors are also considerably reduced.}
    \label{fig:cmp_mri_km22}
\end{figure}

Closest to our work are the universal relations by~\cite{2022ApJ...934..139K}, who consider the increase in mass and radius along a sequence of stars with constant central density; in contrast to the present work, their universal relations are based on the pair $(M_\star, R_\star)$ alone, not including the moment of inertia $I_\star$. For all rotating neutron star models in our dataset, we compute the gravitational mass $M$ and equatorial radius $R_e$ out of the properties of the corresponding non-rotating model. We show histograms of the absolute relative errors according to Eq.~\eqref{eq:re_k} in Fig.~\ref{fig:cmp_mri_km22}. The estimates based on the present work (solid lines) outperform those by~\cite{2022ApJ...934..139K} (dotted lines); in the case of the gravitational mass, the improvement by including $I_\star$ in the estimate is significant. The average magnitude of the relative error decreases from $0.75\%$ to $0.41\%$ for the radius, and from $1.8\%$ to $0.25\%$ for the mass. Furthermore, the maximum relative error is considerably reduced for both quantities.

\cite{2019PhRvD..99d3004R} constructed equation-of-state-independent relations for the Keplerian sequence, expressing the mass-shedding frequency, mass, and radius in terms of the mass and radius of the non-rotating star of the same central density, and improving on the Keplerian-frequency relation of \cite{1990ApJ...355..241L}. Both constructions predict the Keplerian properties from the mass and radius of the reference star alone. Our corresponding relation for the Keplerian angular velocity $\Omega_K$ (cf.~Tab.~\ref{tab:fitting3}) instead employs the moment of inertia $I_\star$ of the reference star in addition to its mass and radius, and estimates $\Omega_K$ with an average discrepancy of $0.14\,\%$ and a worst-case error of $1.0\,\%$. For a direct comparison, we also fit $\Omega_K$ as a function of the compactness $C_\star$ alone, i.e.\ using only $M_\star$ and $R_\star$, as in the works mentioned above as well as \cite{2022ApJ...934..139K}, and obtain an average discrepancy of $0.55\,\%$ and a worst-case error of $3.4\,\%$. Including $I_\star$ thus improves both the average and the worst-case accuracy by more than a factor of three. Since the relations of both Konstantinou \& Morsink and Riahi~\etal\ likewise rely solely on the mass and radius of the reference star, this improvement applies to the class of $(M_\star,R_\star)$-based predictors generally.

\cite{2023PhRvC.108b5810L} likewise relate the integral parameters of maximally rotating stars to those of their non-rotating counterparts. In their case, however, the Keplerian model and the static model to which it is compared share the same baryonic mass, whereas we pair each rotating model with the non-rotating model of the same central density. For the Keplerian spin frequency $f_{\mathrm{K}} = \Omega_K / 2\pi$ they adopt the Newtonian scaling $f_{\mathrm{K}} \propto M^{1/2} R^{-3/2}$, reaching a few per cent accuracy when the mass and radius of the static star are inserted and about $1\,\%$ when those of the Keplerian configuration are used; their relation between the static and Keplerian radii is likewise accurate only at the level of ${\sim}\,10\,\%$. Our relation for $\Omega_K$ supplements $M_\star$ and $R_\star$ by $I_\star$ and attains a worst-case error of $1.0\,\%$ while referring exclusively to the non-rotating configuration; the different choice of the quantity held fixed along the sequence precludes a direct comparison, but the pattern matches the improvement over $(M_\star,R_\star)$-based predictors established above. They further find that the universal relations among the (suitably normalised) moment of inertia, compactness, and quadrupole moment retain comparable accuracy along sequences of constant spin parameter up to the Keplerian one, for equations of state with hyperons and $\Delta$-resonances, independently confirming that the universality survives to the mass-shedding limit.

A complementary line of work extends the $I$-Love-$Q$ relations to the Keplerian limit. \cite{2025PhRvD.112f3028Y}, for example, studied the moment of inertia and quadrupole moment of maximally rotating stars under systematic variation of the nuclear symmetry energy, expressing them as functions of the stellar compactness. In these studies the moment of inertia is a predicted quantity (similar to \cite{2005ApJ...629..979L, 2016MNRAS.459..646B}), obtained from the mass and radius; by contrast, we use the reference-star moment of inertia $I_\star$ as an additional input, and the comparison above shows that it carries information beyond $M_\star$ and $R_\star$.

Universal relations of this kind have also been extended to finite temperature for merger remnants \citep{2020MNRAS.499..914R, 2021PhRvC.103e5811K}, a regime outside the scope of our cold, catalysed models.

\section{Accuracy Tests of Universal Relations}
\label{sec:ur-test}

As a final demonstration of the accuracy of the proposed universal relations for neutron-star bulk quantities, we present the relative errors along sequences of rotating equilibrium models.

In addition to the nine quantities for which we provide universal relations, we also consider quantities that can be estimated from other bulk properties. Derived quantities are indicated by superscripts ``(X)''. For example, instead of using the universal relation for the gravitational binding energy $W$, we may estimate it as
\begin{align}
    W^{(C)} := T / (T/W),
\end{align}
where $T/W$ is obtained from its universal relation and $T = \tfrac{1}{2} I \Omega^2$ is computed from $I$. In this approach the binding energy $W_\star$ of the non-rotating configuration (or, equivalently, the proper mass $M_{p,\star} = M_\star + W_\star$) is not required.

Similarly, the proper mass $M_p$ can be estimated via
\begin{align}
    M_p^{(A)} &= M + W - T, \\
    M_p^{(B)} &= M + W^{(C)} - T.
\end{align}
The expression for $M_p^{(A)}$ follows directly from the definition of the proper mass, while $M_p^{(B)}$ replaces $W$ by $W^{(C)}$ and thus avoids the need for $W_\star$.

We do not discuss the angular momentum $J$ and the rotational kinetic energy $T$ further, as their relative errors are identical to that of the moment of inertia $I$.

To test the universal relations in an astrophysically relevant way, we consider two realistic equations of state available in the CompOSE database \citep{2015PPN....46..633T, 2017RvMP...89a5007O, 2022EPJA...58..221C}. Specifically, we employ the soft APR4 equation of state \citep{1998PhRvC..58.1804A, 2024A&A...687A..44D, 2025EPJA...61..120D, chatterjee_2025_15045814}, which yields a radius of $R_{1.4}=11.37\,\mathrm{km}$ for a $1.4\,M_\odot$ star, and the stiffer DDME2 equation of state \citep{2005PhRvC..71b4312L, 2022CoTPh..74i5303X, 2022PhRvC.105d5803X, 2026PhRvC.113b5804N}, for which $R_{1.4}=13.20\,\mathrm{km}$. Both models include a consistently treated crust. For each equation of state, we construct two constant-central-density sequences whose non-rotating members have gravitational masses of $1.4\,M_\odot$ and $2.0\,M_\odot$. Along each sequence, we compute models with axis ratios of $0.99$, $0.97$, and from $0.95$ down to $0.60$ in steps of $0.025$.

\begin{figure}
    \centering
    \includegraphics[width=1\linewidth]{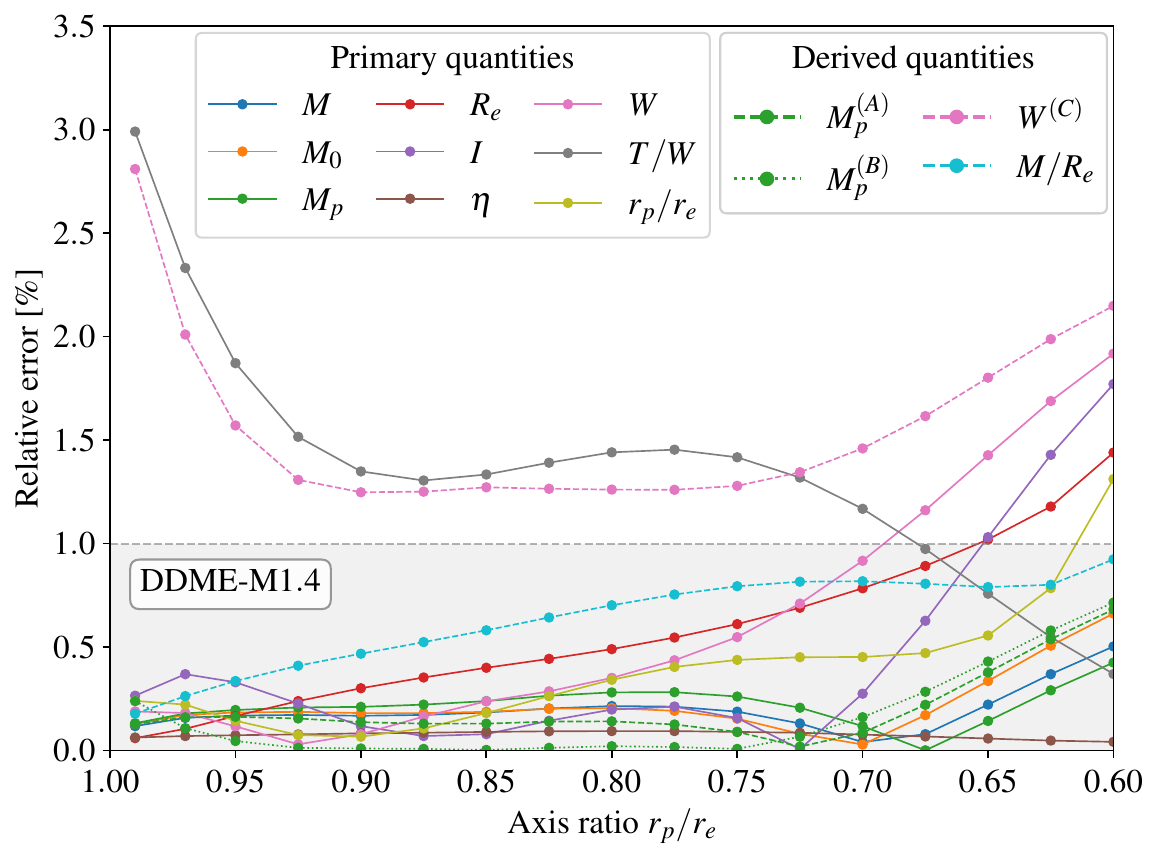}
    \caption{Relative error of the bulk quantities along a sequence of neutron-star models with increasing  rotation. Quantities computed directly from the universal relations are shown with solid lines, whereas derived quantities are indicated by dashed and dotted lines. Except for $T/W$   at low rotation rates and a few quantities ($W$, $I$, $R_e$, $r_p/r_e$) near the Kepler limit, the errors are well below 1 per cent.}
    \label{fig:ddme_1.4}
\end{figure}

\begin{figure}
    \centering
    \includegraphics[width=1\linewidth]{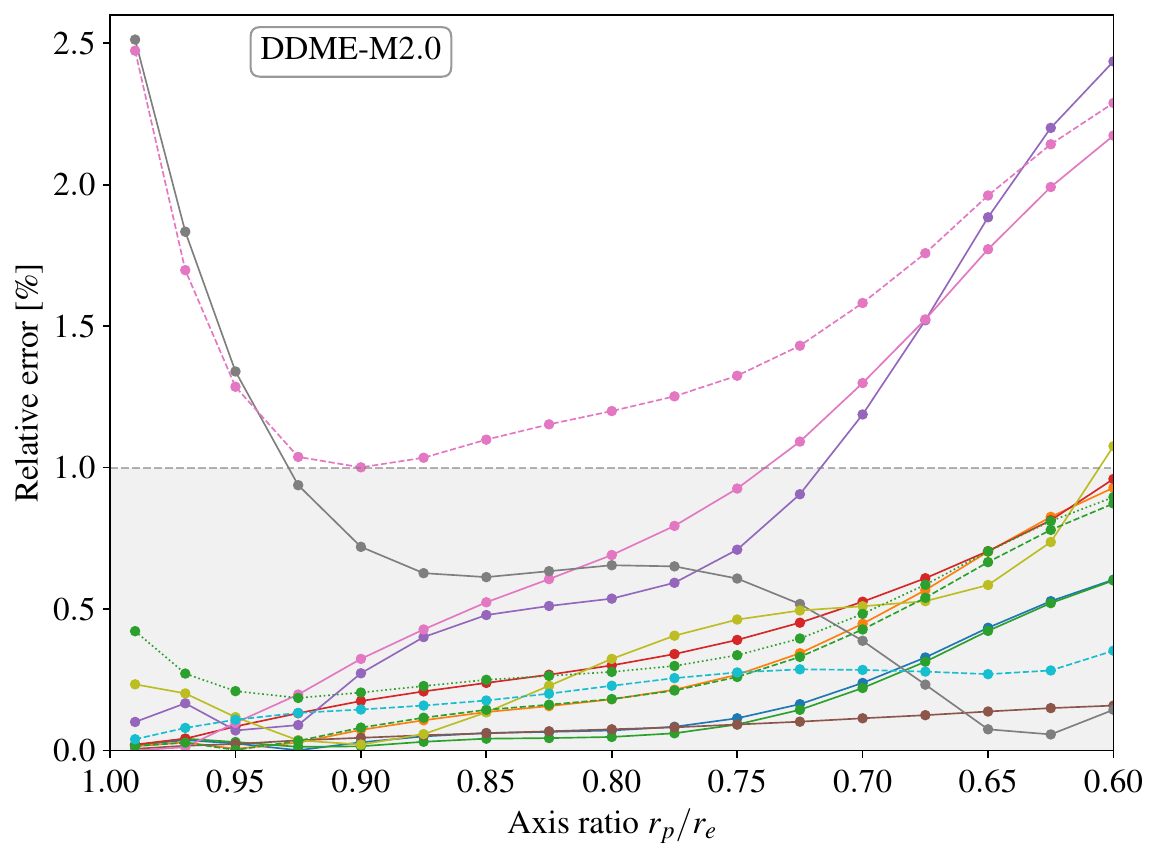}
    \caption{Same as Fig.~\ref{fig:ddme_1.4}, but for the sequence with $M_\star = 2.0\,M_\odot$ using the DDME EOS; the legend is identical.}
    \label{fig:ddme_2.0}
\end{figure}

\begin{figure}
    \centering
    \includegraphics[width=1\linewidth]{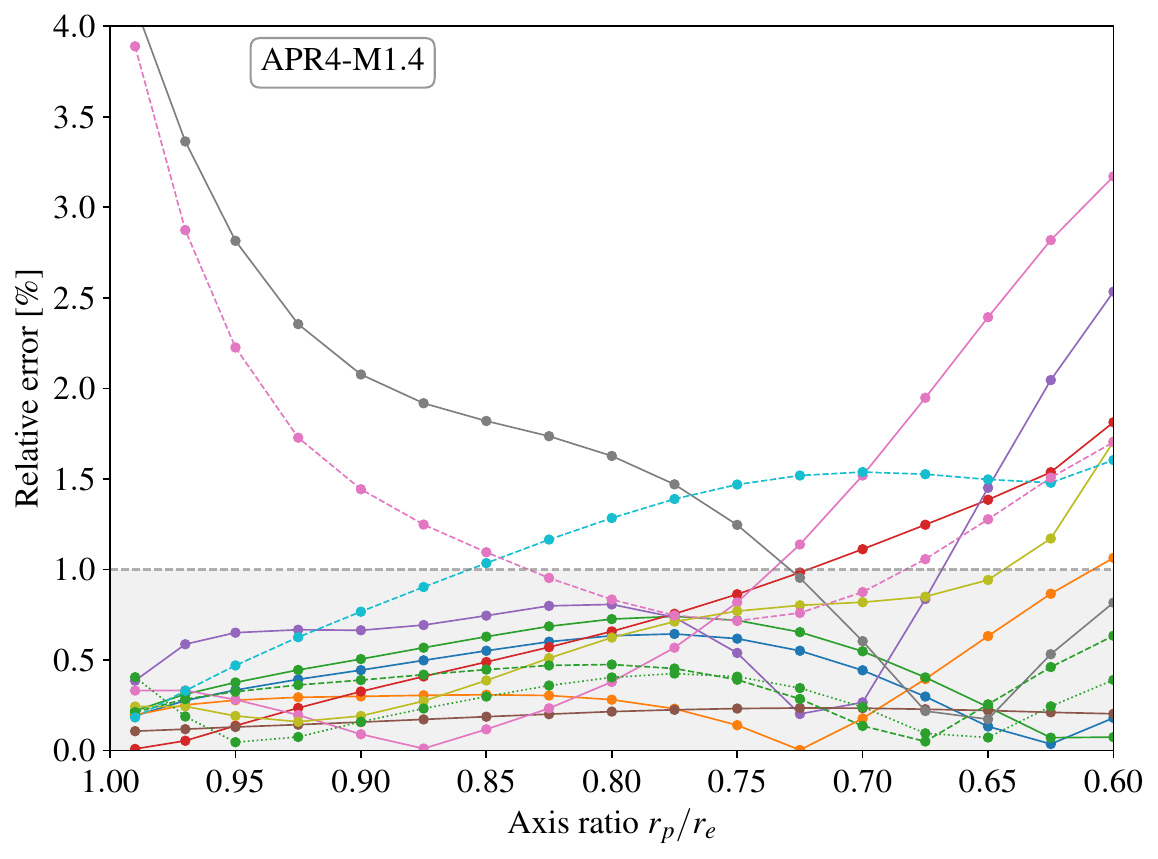}
    \caption{Same as Fig.~\ref{fig:ddme_1.4}, but for the sequence with $M_\star = 1.4\,M_\odot$ using the APR4 EOS; the legend is identical.}
    \label{fig:apr4_1.4}
\end{figure}

\begin{figure}
    \centering
    \includegraphics[width=1\linewidth]{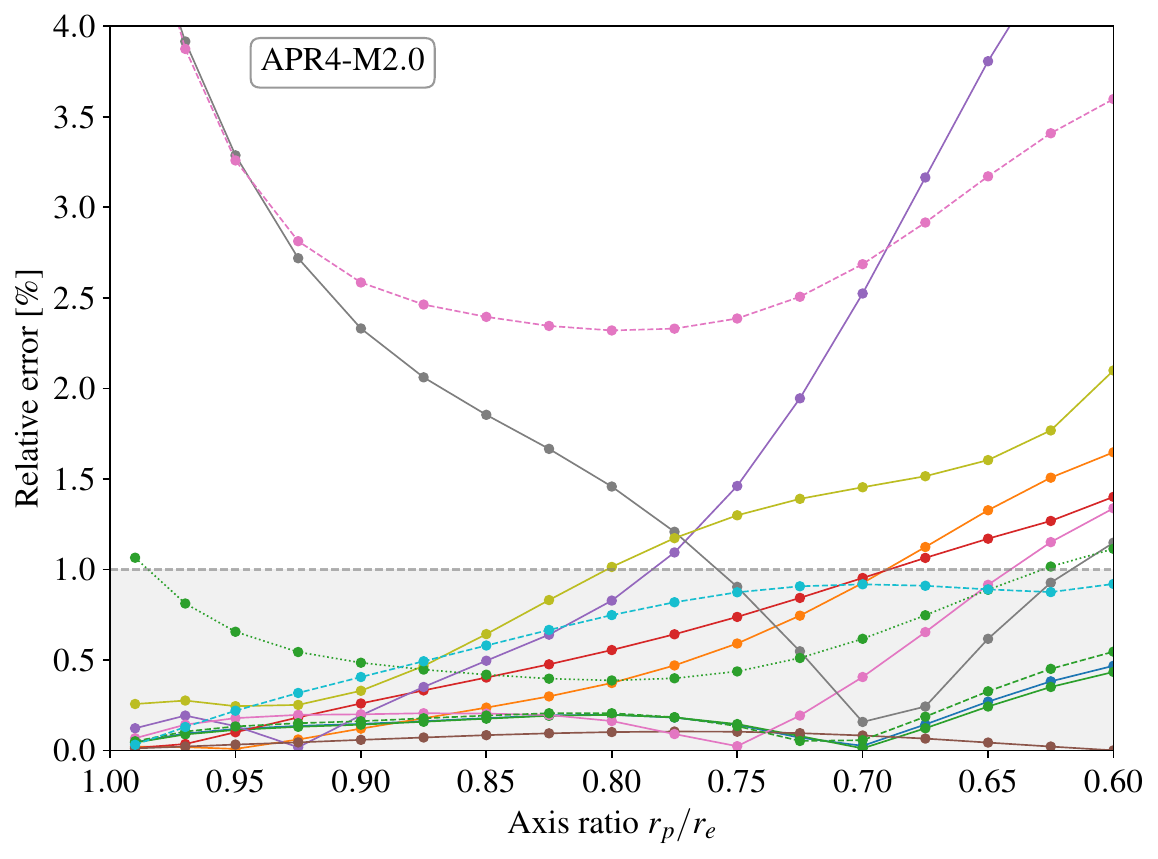}
    \caption{Same as Fig.~\ref{fig:ddme_1.4}, but for the sequence with $M_\star = 2.0\,M_\odot$ using the APR4 EOS; the legend is identical.}
    \label{fig:apr4_2.0}
\end{figure}

For every model we evaluate the relative deviation between the universal-relation estimate and the reference value obtained from the \texttt{rns} code for the nine primary quantities $M$, $M_0$, $M_p$, $R_e$, $I$, $\eta$, $W$, $T/W$, and $r_p/r_e$. We also include the combined quantity $M/R_e$ and the derived quantities $M_p^{(A)}$, $M_p^{(B)}$, and $W^{(C)}$ in our accuracy test. The results are shown in \Cref{fig:ddme_1.4,fig:ddme_2.0,fig:apr4_1.4,fig:apr4_2.0}. We show the dynamical ranges of the bulk quantities exemplarily for the $M = 1.4\,M_\odot$ DDME2 sequence explicitly in Tab.~\ref{tab:ddme14} in order to demonstrate that the relative error of the universal relations is drastically lower than the relative change of the quantities themselves.

For most quantities the relative errors remain well below $1\,\%$ over the entire rotation range. A notable exception is $T/W$, which exhibits large relative errors at low rotation rates because the quantity vanishes in the non-rotating limit, so that even small absolute errors lead to large relative deviations. The accuracy of the moment of inertia $I$ also deteriorates for axis ratios below $\approx 0.7$, and the derived quantity $W^{(C)}$ inherits this behaviour as it depends on both $I$ and $T/W$. Interestingly, in some cases (e.g.\ APR4 with $M_\star = 1.4\,M_\odot$) the alternative estimate $W^{(C)}$ provides a more accurate value for the binding energy near the Kepler limit than the direct universal-relation estimate for $W$.

The three estimates for the proper mass $M_p$ show comparable accuracy throughout most of the sequence. Only $M_p^{(B)}$ performs worse at low rotation rates due to its reliance on $W^{(C)}$, partially offsetting the advantage of not requiring $M_{p,\star}$.

\begin{table}
    \centering
    \caption{Variation of the primary and derived quantities between the non-rotating configuration and that with $\mathfrak{r} = 0.6$, together with the relative change of the quantities over that range as well as the maximal relative error of the universal relations for the DDME2 sequence with $M_\star = 1.4\,M_\odot$ (Fig.~\ref{fig:ddme_1.4}); the fractional change of the bulk quantities is comparable for the other sequences. The maximal relative error is usually attained close to the Kepler limit, the exception being $T/W$ and $W^{(C)}$, which show large discrepancies at low rotation rates due to $T \to 0$.}
    \sisetup{
        table-number-alignment=center,
        detect-all
    }
    \begin{tabular}{l|
        S[table-format=+2.3]
        S[table-format=+2.3]
        S[table-format=+2,print-implicit-plus]
        S[table-format=+1.2]}
    \toprule
    {Quantity} & {Non-rot.} & {$\mathfrak{r} = 0.6$} & {Change} & {Max RE} \\
             &          &        & [\%]   & [\%] \\
    \midrule
    $M\,[M_\odot]$                   & 1.40  & 1.80  & +29 & 0.50 \\
    $M_0\,[M_\odot]$                 & 1.53  & 1.98  & +29 & 0.66 \\
    $M_p\,[M_\odot]$                 & 1.60  & 2.06  & +29 & 0.43 \\
    $R_e\,[\mathrm{km}]$             & 13.18 & 17.44 & +32 & 1.44 \\
    $I\,[10^{45}\,\unit{g.cm^2}]$    & 1.72  & 3.30  & +92 & 1.77 \\
    $\eta$                           & 0.263 & 0.277 & +5  & 0.10 \\
    $W\,[M_\odot]$                   & 0.194 & 0.292 & +51 & 1.92 \\
    $T/W$                            & 0.000 & 0.116 & {---} & 3.00 \\
    $r_p/r_e$                        & 1.00  & 0.60  & -40 & 1.31 \\
    \midrule
    $M_p^{(A)}\,[M_\odot]$           & 1.60  & 2.06  & +29 & 0.69 \\
    $M_p^{(B)}\,[M_\odot]$           & 1.60  & 2.06  & +29 & 0.72 \\
    $W^{(C)}\,[M_\odot]$             & 0.194 & 0.292 & +50 & 2.81 \\
    $M/R_e$                          & 0.157 & 0.152 & -3  & 0.93 \\
    \bottomrule
    \end{tabular}
    \label{tab:ddme14}
\end{table}

We emphasise that the equations of state used for testing include a consistently modelled crust, in contrast to the simplified crust treatment adopted in the construction of the universal relations. The good agreement obtained for the equatorial radius $R_e$ therefore demonstrates the robustness of the relation against variations in the crust modelling.

\section{Summary}
\label{sec:summary}

In the first part of the paper, we have presented nine different equation-of-state parametrisations that have been proposed during the past two decades and parametrise our ignorance of the high density part of the nuclear equation of state of neutron stars. The parametrisations have been developed for different purposes such as simplicity over tabulated equations of state, for studies on the speed-of-sound profile in the cores of neutron stars, with the intent of improved smoothness in order to increase the convergence rate in binary simulations, or others. We discuss the various parameters that are introduced in these parametrisations and list reasonable ranges within which these should lie in order to generate ``mostly reasonable'' equations of state; further, we mention correlations between these parameters that are present in some of the parametrisations.

Not all of the generated equations of state satisfy astrophysical bounds, and some of them even violate fundamental physical laws such as causality. We impose generous astrophysical constraints based on prior observations and condense our data set to 320 equations of state for each parametrisation that fulfill these constraints. For each equation of state, we construct several sequences of stars with the same central energy density with rotation rates from no rotation up to the mass-shedding limit; our data set contains 21383 sequences with 998639 neutron star models in total. We also test a data set based on more recent, considerably tighter astrophysical constraints but the accuracy of the results are improved only marginally.

Based on this large data set, we propose several universal relations that allow to approximate various bulk quantities of uniformly rotating neutron stars to percent-level accuracy at minimal computational cost irrespective of the parametrisation or the stellar rotation rate; in our tests, the speed-up compared to solving the full equilibrium equations with the \textsc{rns} code is roughly three orders of magnitude for the computationally very simple piecewise polytropes and accordingly higher for the more complex parametrisations. In particular, we propose universal relations to obtain the Keplerian values of bulk quantities and subsequently also for rotating neutron stars at arbitrary rotation rates. These universal relations are insensitive to the equation of state (and in particular insensitive to the parametrisations which may introduce a bias by themselves) and rely solely on the knowledge of the triple $(M_\star, R_\star, I_\star)$ of the non-rotating star with the same central energy density, which can easily and computationally cheaply be calculated by solving the TOV equations along with Hartle's equation.

The presented universal relations show a high level of accuracy, despite being constructed based on randomly generated equations of state that satisfy some generous astrophysical bounds rather than on equations of state that are microphysically or phenomenologically motivated; they allow to estimate bulk quantities of rotating stars at low computational cost. We compare the performance of our relations to previously published relations and find that they yield significantly more accurate results; we attribute this improvement to the additional information contained in $I_\star$. Such universal relations facilitate cheap equation of state inference codes and may also be used in gravitational wave modelling.

Tests with tabulated equations of state including a realistically modeled crust, for sequences with $M_\star = 1.4\,M_\odot$ and $2.0\,M_\odot$ up to the Kepler limit, show that the universal relations reproduce bulk quantities with typical relative errors below 1\%. Derived quantities perform similarly well, with somewhat increased deviations for $T/W$ at low rotation and a few quantities near the Kepler limit, confirming the robustness of the proposed relations.

\section*{Acknowledgements}
The authors are grateful to Erich Gaertig and Sebastian V\"olkel for very valuable discussions and insightful comments on the manuscript. They also thank Christian Eckert, Tyler Gorda, Kostas Kokkotas, and Achim St\"o\ss{}l for useful comments and helpful feedback. MC contributed the implementation of the various spectral representations and corresponding tests. The authors thank two anonymous referees for valuable comments and suggestions that helped improve this manuscript. The authors acknowledge support by the High Performance and Cloud Computing Group at the Zentrum f\"ur Datenverarbeitung of the University of T\"ubingen, the state of Baden-W\"urttemberg through bwHPC and the German Research Foundation (DFG) for funding under ``Project number 455787709'' (bwForCluster BinAC 2).

\section{Data availability}
The data underlying this article will be shared on reasonable request to the corresponding author.

\appendix

\section{$M$-$R$ diagrams for the different parametrisations}
\label{app:m-r}

These figures are provided for completeness and to illustrate the qualitative differences between the parametrisations discussed in Sec.~\ref{sec:eos_params}.

\begin{figure}
    \centering
    \includegraphics[width=1\linewidth]{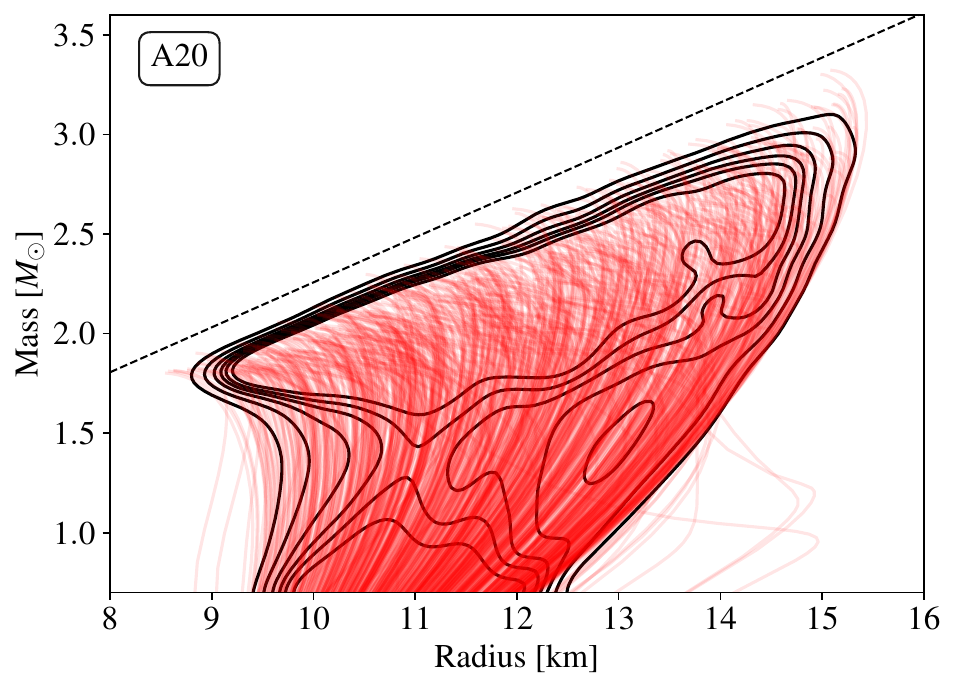}
    \caption{Mass–radius curves (red) for all 1000 randomly generated equations of state based on parametrisation A20. Black contour lines indicate the KDE of the data at the 50\%, 60\%, 70\%, 80\%, 90\%, and 95\% levels. The black dashed line shows the compactness $C = 1/3$.}
    \label{fig:m-r-a20}
\end{figure}

\begin{figure}
    \centering
    \includegraphics[width=1\linewidth]{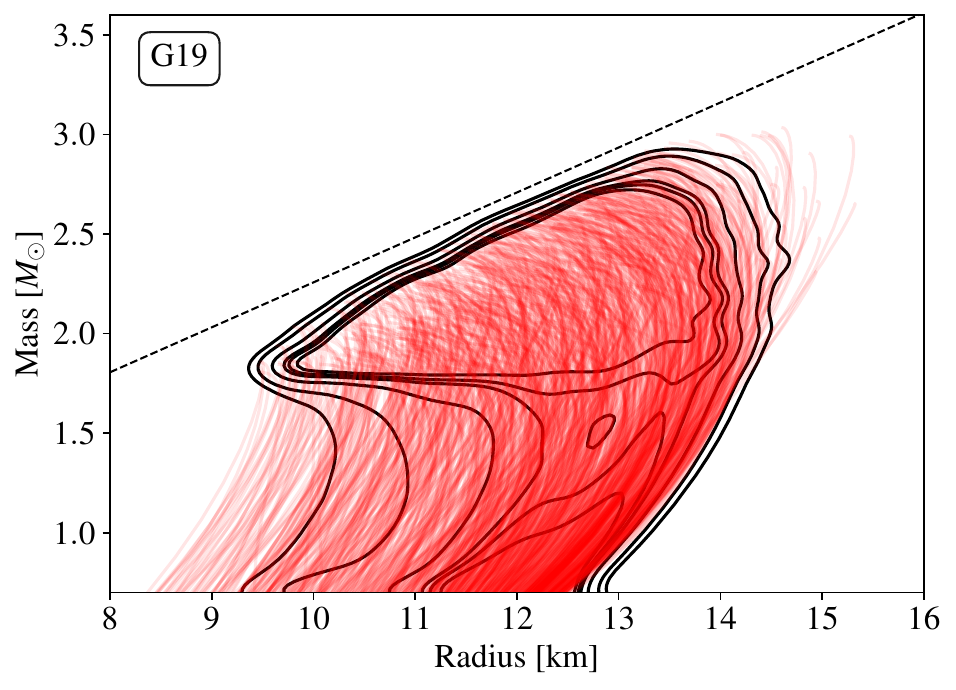}
    \caption{Same as Fig.~\ref{fig:m-r-a20} but for the parametrisation G19.}
    \label{fig:m-r-g19}
\end{figure}

\begin{figure}
    \centering
    \includegraphics[width=1\linewidth]{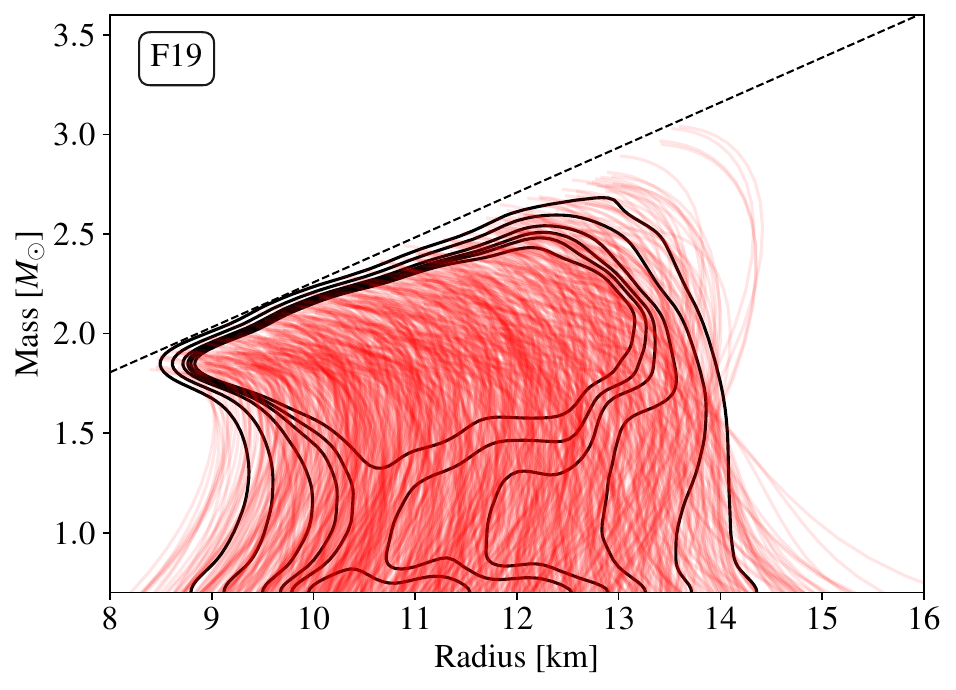}
    \caption{Same as Fig.~\ref{fig:m-r-a20} but for the parametrisation F19.}
    \label{fig:m-r-f19}
\end{figure}

\begin{figure}
    \centering
    \includegraphics[width=1\linewidth]{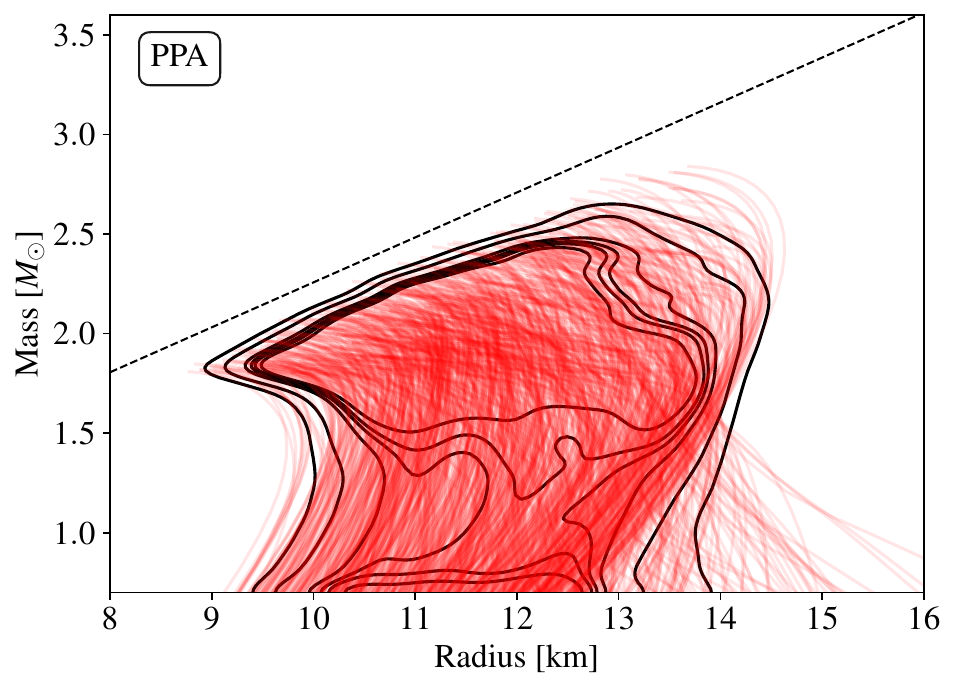}
    \caption{Same as Fig.~\ref{fig:m-r-a20} but for the parametrisation PPA.}
    \label{fig:m-r-ppa}
\end{figure}

\begin{figure}
    \centering
    \includegraphics[width=1\linewidth]{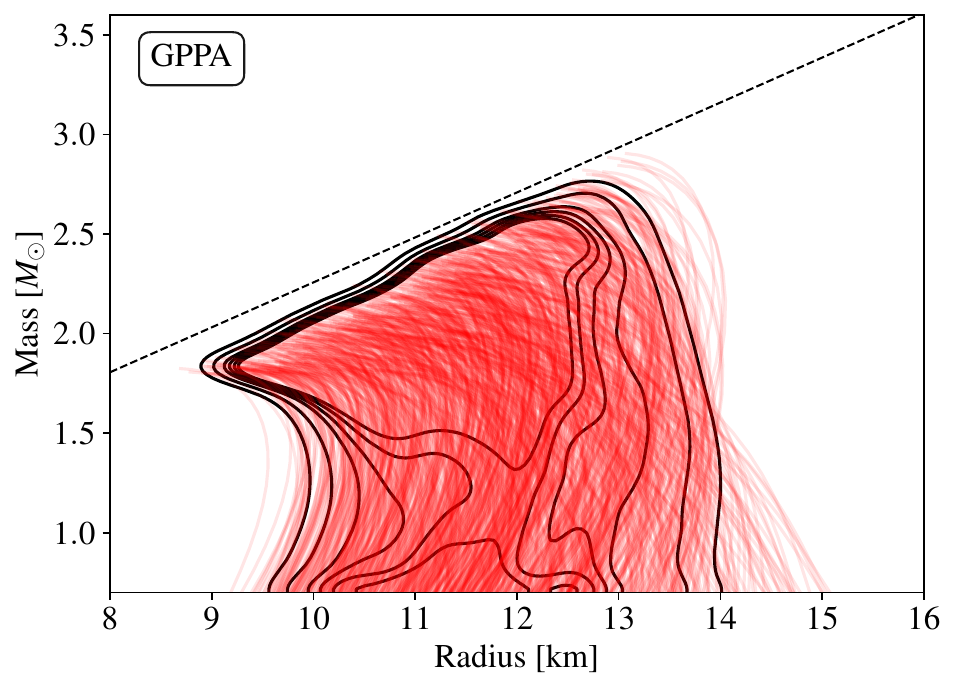}
    \caption{Same as Fig.~\ref{fig:m-r-a20} but for the parametrisation GPPA.}
    \label{fig:m-r-gppa}
\end{figure}

\begin{figure}
    \centering
    \includegraphics[width=1\linewidth]{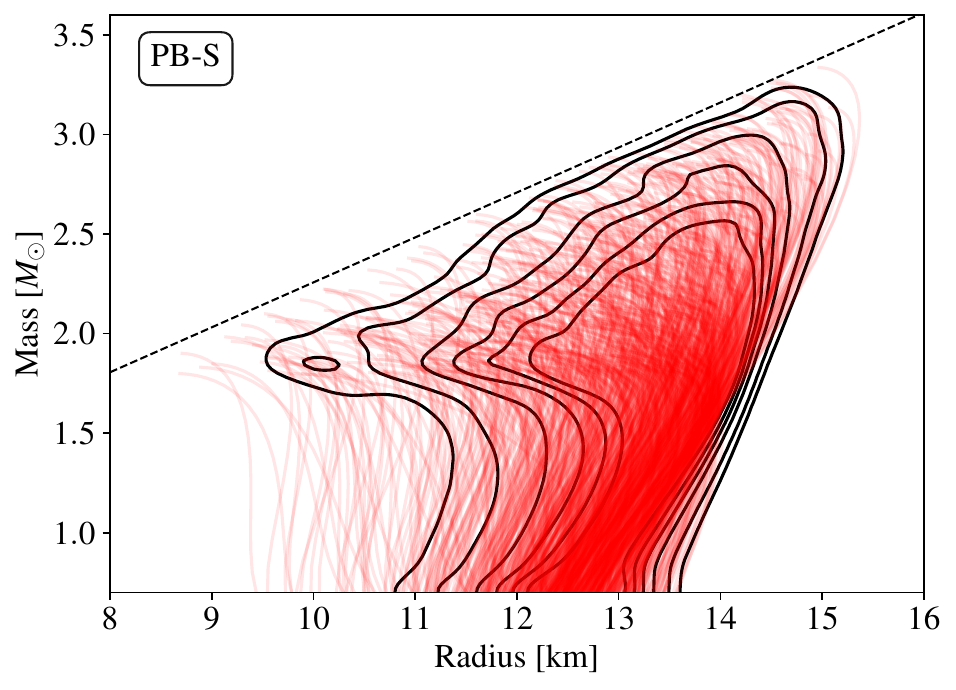}
    \caption{Same as Fig.~\ref{fig:m-r-a20} but for the parametrisation PB-S.}
    \label{fig:m-r-pb-s}
\end{figure}

\begin{figure}
    \centering
    \includegraphics[width=1\linewidth]{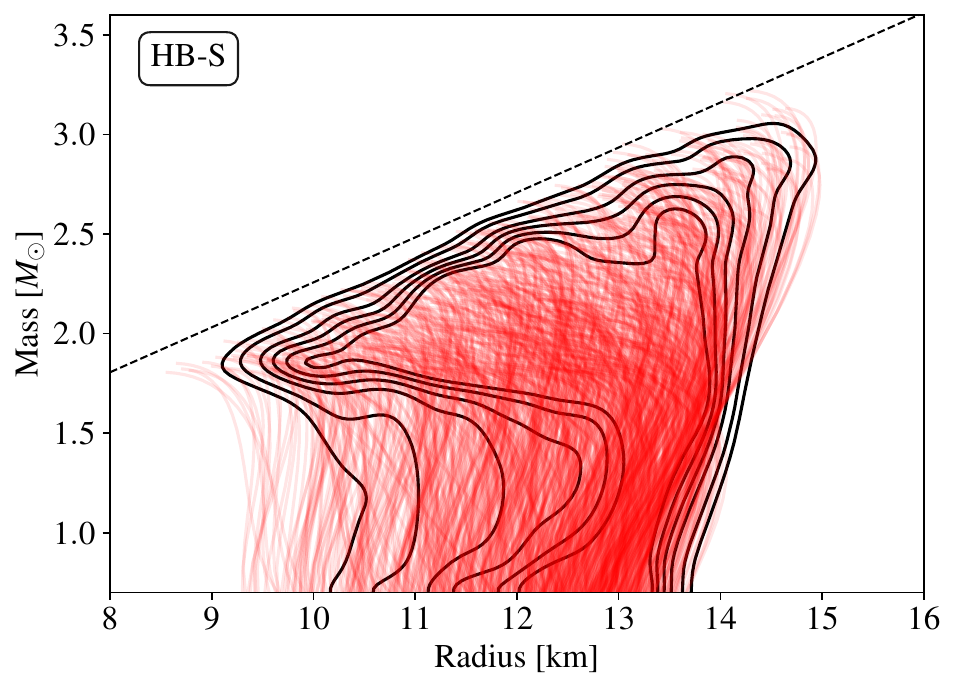}
    \caption{Same as Fig.~\ref{fig:m-r-a20} but for the parametrisation HB-S.}
    \label{fig:m-r-hb-s}
\end{figure}

\begin{figure}
    \centering
    \includegraphics[width=1\linewidth]{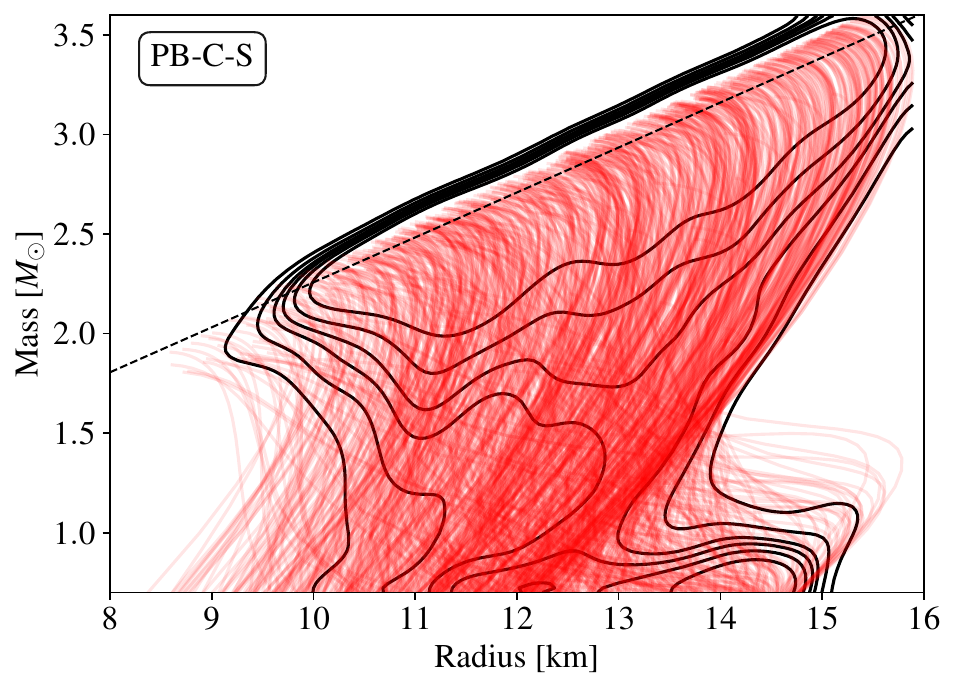}
    \caption{Same as Fig.~\ref{fig:m-r-a20} but for the parametrisation PB-C-S.}
    \label{fig:m-r-pb-c-s}
\end{figure}

\begin{figure}
    \centering
    \includegraphics[width=1\linewidth]{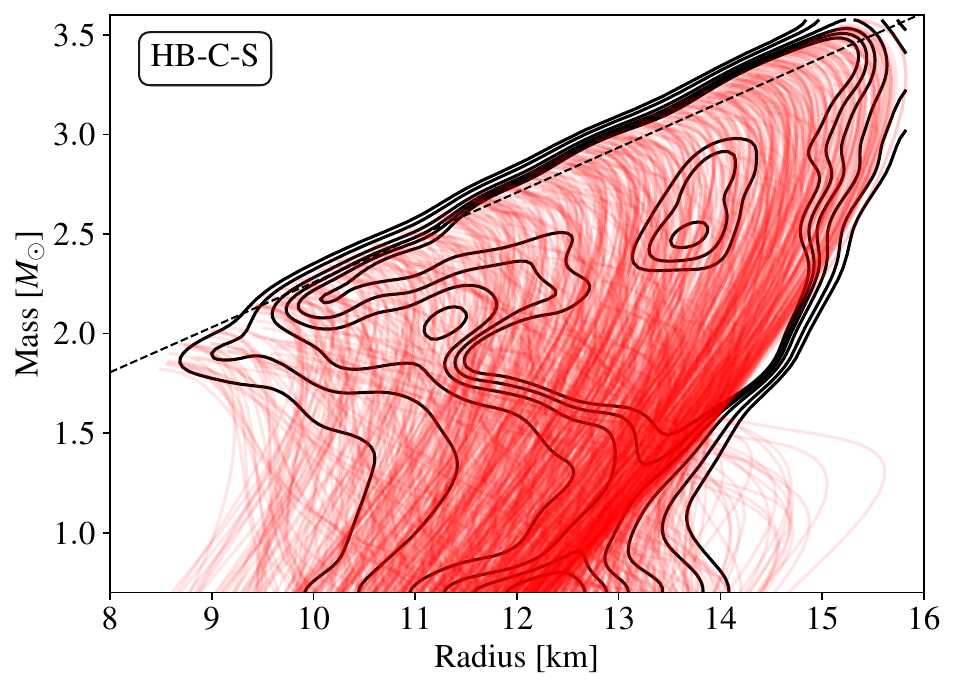}
    \caption{Same as Fig.~\ref{fig:m-r-a20} but for the parametrisation HB-C-S.}
    \label{fig:m-r-hb-c-s}
\end{figure}

We show the non-rotating mass--radius diagrams of the nine employed parametrisations, including all 1000 randomly generated equations of state, in~\Cref{fig:m-r-a20,fig:m-r-g19,fig:m-r-f19,fig:m-r-ppa,fig:m-r-gppa,fig:m-r-pb-s,fig:m-r-hb-s,fig:m-r-pb-c-s,fig:m-r-hb-c-s}. The data shown are subject only to the basic filtering described in the first paragraph of Sec.~\ref{sec:dataset}. To clarify the density of models in the diagrams, we overlay kernel density estimates (KDEs) at six different confidence levels. In addition, we indicate the line corresponding to a compactness $C = GM/(Rc^2) = 1/3$, which was empirically identified by \cite{2026ScPP...20..014R} as an upper bound for neutron star compactness. The maximum compactness values (attained at the maximum-mass configuration) observed in our dataset are summarised in Tab.~\ref{tab:cmax_parametrisations}. Notably, almost all parametrisations respect this bound. The only exceptions are the spectral representations: while the PB-S and HB-S parametrisations show mild violations, the speed-of-sound--based parametrisations PB-C-S and HB-C-S more frequently produce equations of state that exceed $C=1/3$ and reach compactness values of up to \(C \approx 0.346\).

A qualitative comparison of the mass--radius curves shows that the different parametrisations tend to produce equations of state with varying characteristics. In particular, the G19, F19, PPA, and GPPA parametrisations only rarely support neutron stars with masses around \(3\,M_\odot\), whereas the remaining parametrisations more readily reach such masses or beyond. Conversely, the F19, PPA, and GPPA parametrisations typically allow for larger stellar radii at lower masses. We note, however, that we have imposed no upper constraint on the radius of a \(1.4\,M_\odot\) neutron star; applying such a constraint would likely modify these trends.

\begin{table}
\centering
\caption{Maximum compactness $C_{\max}$ for the different equation-of-state parametrisations considered.}
\begin{tabular}{l c}
\toprule
Parametrisation & $C_{\max}$ \\
\midrule
A20     & 0.3262 \\
G19     & 0.3270 \\
F19     & 0.3328 \\
PPA     & 0.3242 \\
GPPA    & 0.3325 \\
PB-S    & 0.3352 \\
HB-S    & 0.3382 \\
PB-C-S  & 0.3463 \\
HB-C-S  & 0.3460 \\
\bottomrule
\end{tabular}
\label{tab:cmax_parametrisations}
\end{table}

\end{document}